\documentclass[%
 aip,
 amsmath,amssymb,
 reprint,
 floatfix,%
]{revtex4-1}

\usepackage{graphicx}
\usepackage{dcolumn}
\usepackage{bm}

\usepackage[utf8]{inputenc}
\usepackage[T1]{fontenc}
\usepackage{mathptmx}
\usepackage{xcolor}

\defcitealias{Acrivos1993}{Acrivos {\it et al.}}
\defcitealias{Zarraga2000}{Zarraga {\it et al.}}
\defcitealias{Boyer2011}{Boyer {\it et al.}}

\begin{document}

\preprint{APS/123-QED}

\title{X-ray radiography of viscous resuspension}

\author{Brice Saint-Michel}
\affiliation{Department of Chemical Engineering, Imperial College London, London SW7 2AZ, United Kingdom}
\affiliation{Univ  Lyon,  Ens  de  Lyon,  Univ  Claude  Bernard, CNRS,  Laboratoire  de  Physique,  F-69342  Lyon,  France}%
\email{b.saint-michel@imperial.ac.uk}

\author{S\'ebastien Manneville}
\affiliation{Univ  Lyon,  Ens  de  Lyon,  Univ  Claude  Bernard, CNRS,  Laboratoire  de  Physique,  F-69342  Lyon,  France}%

\author{Steven Meeker}%
\affiliation{Univ.  Bordeaux, CNRS, Solvay, LOF, UMR 5258,  F-33608  Pessac,  France}%

\author{Guillaume Ovarlez}
\affiliation{Univ.  Bordeaux, CNRS, Solvay, LOF, UMR 5258,  F-33608  Pessac,  France}%

\author{Hugues Bodiguel}
\affiliation{Univ.  Grenoble  Alpes, CNRS, Grenoble-INP, Laboratoire  Rh\'eologie  et  Proc\'ed\'es, UMR 5520, F-38041  Grenoble,  France}%

\date{\today}

\begin{abstract}
We use X-ray imaging to study viscous resuspension. In a Taylor-Couette geometry, we shear an initially settled layer of spherical glass particles immersed in a Newtonian fluid and measure the local volume fraction profiles. In this configuration, the steady-state profiles are simply related to the normal viscosity defined in the framework of the Suspension Balance Model (SBM). These experiments allow us to examine this fundamental quantity over a wide range of volume fractions, in particular in the semi-dilute regime where experimental data are sorely lacking. Our measurements unambiguously show that the particle stress is quadratic with respect to the volume fraction in the dilute limit. Strikingly, they also reveal a nonlinear dependence on the Shields number, in contrast with previous theoretical and experimental results. This likely points to shear-thinning particle stresses and to a non-Coulomb or velocity-weakening friction between the particles, as also evidenced from shear reversal experiments.
\end{abstract}

\maketitle

\section{Introduction}

Understanding the flow of particles suspended in a fluid is critical to obtaining reliable predictions and models of transport and migration phenomena in industrial and natural slurries. This problem has attracted significant attention over the last two centuries, starting from the seminal works of \citet{Stokes1851} and \citet{Einstein1906} who studied how a single particle affects the flow of a viscous fluid at low Reynolds number. Due to the large number of particles in the fluid and to the presence of both solid contacts and long-range hydrodynamic interactions between them, addressing the full problem of suspension flows remains a complex challenge even after almost a hundred years of theoretical and experimental effort. The last two decades have seen the emergence of the suspension balance model (SBM)~\cite{Nott1994,Morris1999} as a robust framework to describe migration phenomena in suspensions, despite some conceptual flaws that were settled recently~\cite{Lhuillier2009,Nott2011}. Notable progress has also been made to reconcile this theoretical framework for dense suspensions with those for granular rheology~\cite{Boyer2011}. 

The SBM introduces the concept of particle stress to explain particle migration in flowing suspensions. It accounts for the fact that particle contacts tend to generate normal stresses that act as an osmotic pressure to keep the particles dispersed under shear. The normal viscosity tensor $\eta_n$ quantifies this particle stress in a dimensionless form as a function of the volume fraction $\phi$. In the presence of stress inhomogeneities, particle migration occurs leading to volume fraction gradients that ensure the balance of particle normal stresses in steady-state conditions.

The SBM then relies on empirical expressions of the shear and normal viscosities of dense suspensions derived from experiments performed under homogeneous conditions, such as those provided by \citet{Morris1999} and \citeauthor{Boyer2011}\cite{Boyer2011,Guazzelli2018}, to derive the particle migration dynamics and the steady-state concentration profiles in any flow geometry~\cite{Morris1999}. In all these works, a viscous scaling of all stresses is assumed, which is theoretically justified for ideal rate-independent Coulomb friction between the particles~\cite{Guazzelli2012}. Many experimental results, however, show the emergence of a shear-thinning viscosity at high volume fraction in non-Brownian suspensions~\cite{Zarraga2000,Dbouk2013,Chatte2018,Tanner2018, Lobry2019}. This has been proposed to originate from a velocity-weakening friction between the particles~\cite{Tanner2018} or from non-Coulomb friction~\cite{Chatte2018,Lobry2019} as evidenced experimentally by~\citet{Chatte2018}. These nonlinear properties would then largely depend on both the bulk and surface properties of the particles considered.

Moreover, some aspects of the SBM remain unclear. For instance, the anisotropy in the particle pressure is modelled through the use of three different coefficients for the normal viscosity, $\eta_{n,i=1,2,3}$ acting respectively in the direction of flow, in the flow gradient direction and in the vorticity direction, perpendicular to both flow and flow gradient directions. \citet{Morris1999} assume that these coefficients are proportional to each other for all $\phi$. However, recent experimental evidence by~\citet{Dbouk2013}, who measured the particle pressure simultaneously in two directions, have shown otherwise.  Confirming such results is particularly challenging because very few experimental configurations probe the particle pressure in the vorticity direction. Such measurements indeed require a parallel-plate geometry~\cite{Dbouk2013,Gamonpilas2016} or a tilted trough~\cite{Couturier2011}. The typical particle pressures in these geometries are usually too small to be measured accurately for $\phi \leq 0.2$, which means that the asymptotic limits proposed by \citet{Morris1999} have yet to be confirmed. Additional measurements of the normal viscosity in the vorticity direction would allow one to confirm such predictions or to correct them and improve our understanding of particle migration and transport in dilute and semi-dilute suspensions.

Viscous resuspension~\cite{Leighton1986,Acrivos1993}, where an initially settled suspension is made to flow until a steady state is reached, is an interesting alternative to measure directly the normal viscosity coefficients. Indeed, under a homogeneous and steady shear flow, particle pressure gradients originating from the initial particle concentration inhomogeneity induce a vertical positive particle flux --that is, particles are resuspended-- which is balanced in steady state by the negative buoyancy of the particles. Resuspension experiments have been performed in various geometries including the annulus geometry~\cite{Leighton1986} and the Taylor-Couette geometry~\cite{Acrivos1993}. Due to technical limitations, these two previous works only report measurements of the maximum height of the sediment $h$ above which the particle volume fraction is identically zero. This height increases when raising the shear rate $\dot{\gamma}$, as expected from the balance between viscous forces and buoyancy. More specifically, in the context of viscous resuspension in a Taylor-Couette geometry, the increment $h-h_0$ of an initially settled bed can be related to the normal viscosity in the vorticity direction. Interestingly, the data of \citetalias{Acrivos1993}, analysed in the framework of the suspension balance model by \citet{Zarraga2000}, suggest an alternative dependence of the normal viscosity with $\phi$ compared to pressure-imposed experiments~\cite{Morris1999,Boyer2011}. In particular, for vanishing volume fractions $\phi$, this alternative expression converges faster to zero than the ones proposed by \citet{Morris1999} and by \citetalias{Boyer2011} (see Section~\ref{sec:theory} for more details). 

The height of the sediment in resuspension experiments is an integral response, and as such it includes both the dilute particle layers close to the top of the sediment and the more concentrated parts at the bottom. However, in the steady state, a mechanical balance links the vertical concentration profile to the normal viscosity everywhere in the sediment, or, in other words, at every concentration from the very dilute regime to the highly concentrated regime. Therefore, concentration profiles in Taylor-Couette resuspension experiments could be used to estimate the normal viscosity for a wide range of volume fractions, including the very dilute limit. This would be particularly helpful to reconcile the point of view of \citetalias{Zarraga2000} and that of \citet{Morris1999} and \citetalias{Boyer2011} on the expression of the normal viscosity.

The objective of the present work is to achieve the measurement of volume fraction profiles $\phi(z)$ during viscous resuspension in a Taylor-Couette geometry and to use this information to determine the normal viscosity. For this purpose we take advantage of X-ray imaging, since X-ray absorbance is directly related to the particle volume fraction without being affected by multiple scattering~\cite{Chatte2018,Deboeuf2018,Gholami2018}. This technique allows a precise and local measurement of the particle volume fraction. The  paper is organized as follows. Section~\ref{sec:setup} gives a detailed description of the experimental setup and the derivation of the concentration profiles from the raw X-ray radiography intensity maps. Section~\ref{sec:observations} discusses our experimental concentration maps and vertical profiles. Our experimental profiles are then compared to the predictions of the SBM using the two empirical formula for the normal viscosity~\cite{Zarraga2000,Boyer2011} in Section~\ref{sec:compare}. Finally, Section~\ref{sec:discussion} offers an interpretation of our results in the framework of solid friction weakening and shear-thinning.

\section{Experimental Setup and Methods}
\label{sec:setup}

\begin{figure}[h]
    \centering
    \includegraphics[scale = 0.68] {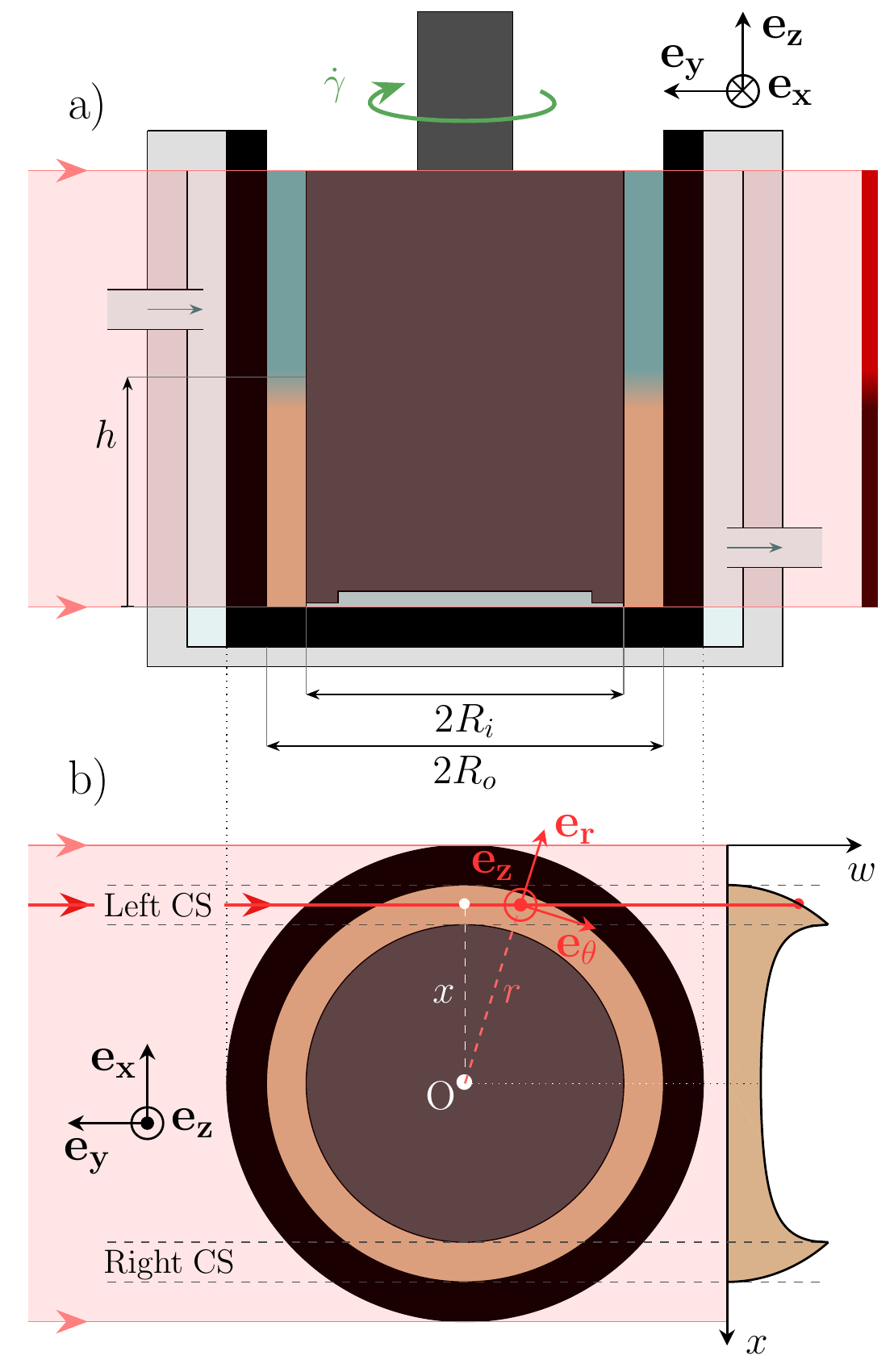}
    \caption{Schematic diagram of the experimental setup. a) Side view: the partially settled suspension (fluid in blue and particles in orange) is sheared between the two concentric cylinders of a Taylor-Couette cell (inner spindle in dark gray and outer cup in black). X-rays generated from a pointlike source 25~cm from the geometry (left of the picture, not shown) are selectively absorbed by the suspension, which translates into spatial variations in the transmitted intensity recorded on the planar X-ray detector located 70~cm away from the cell in the actual experiment (thick red line on the right). The sample temperature is controlled thanks to a water circulation (in gray) around the outer cylinder. b) Top view: the thickness $w$ of the suspension slab crossed by the X-rays depends on the transverse coordinate $x$. The two cross-sections (left and right CS) used in our study are indicated with dashed lines and a plot of $w$ as a function of $x$ is shown on the right. A typical X-ray for a given $x$ of the left CS (thick red line) is also shown to highlight that it crosses areas of the geometry corresponding to multiple values of $r \geq x$.}
    \label{fig:setup}
\end{figure}

\subsection{Geometry, Fluid and Particles}

A schematic diagram of our experimental setup is shown in Figure~\ref{fig:setup}. It consists of a Taylor-Couette cell made of an inner cylindrical spindle of radius $R_i = 23.0$~mm and height $H = 53.5$~mm and of an outer cup of radius $R_o = 25.0$~mm. The choice of this thin-gap Couette geometry is prescribed by the need to avoid significant shear-induced migration while accommodating at least $\sim 8$ particles across the gap. Both cylinders are made of poly(methyl-methacrylate) (PMMA) and are smooth compared to the particle typical size. The spindle is recessed over 1.5~mm at its end in order to limit secondary flows at the bottom of the geometry. Water is circulated around the cup to control the temperature to $25 \pm 1^\circ$C. 

We focus on glass spheres of density $\rho_p = 2500$~kg$\times$m$^{-3}$. The batch particles (purchased from Wheelabrator) were sieved leading to a diameter range $250 < 2a < 315~\mu$m. We suspend these particles in a mixture of water at 65\% wt. and UCON (Dow Chemical, 75-H-90,000) at 35\% wt. The suspending fluid mixture is Newtonian with a viscosity $\eta_0\simeq 0.23$~Pa.s$^{-1}$ and a density $\rho_f \simeq 1.03\times 10^3$~kg$\times$m$^{-3}$. The total particle mass $M = 4.00$~g was weighed using a precision scale before inserting them in the geometry and the global particle volume fraction is $\phi=10\%$. 

Assessing the shear viscosity of such suspensions is difficult: strong sedimentation effects result in a suspension that is non-homogeneous in the vertical direction. We can however evaluate this viscosity by conducting additional experiments with suspensions composed of the same particles suspended in a similar but much more viscous fluid phase (20\% wt. water, 80\% wt. UCON), for which homogeneous steady states under shear are \textit{a priori} reached. These additional results are shown in Appendix~\ref{sec:app:viscosity}. They evidence that the shear viscosity follows a standard Maron-Pierce like evolution with volume fraction, which seems to diverge at $\phi_m \simeq 0.6$, in addition to weak shear-thinning at the highest concentrations. We will thus consider that the SBM is \textit{a priori} an appropriate framework to describe these suspensions rheology.

In the resuspension experiments, the spindle is driven by a stress-imposed rheometer (AR G2, TA Instruments). The vertical position of the spindle is set as the lowest position that allows free rotation, usually around 50-100~$\mu$m from the cup bottom, in order to prevent particles from getting below the spindle. We use the rheometer feedback loop on the imposed stress to apply a constant shear rate $\dot\gamma = 1000$, $500$, $250$, $100$, $50$, $25$ and $0$~s$^{-1}$ in successive steps of 5~min duration each.

\subsection{X-ray Radiography}
\label{sec:Xradio}

The whole rheometer and the Taylor-Couette cell are inserted in a high-resolution X-ray apparatus (Phoenix v$\vert{}$tome$\vert{}$x s, GE) set to work in two-dimensional mode with a pointlike source (see Figures~\ref{fig:setup}a, \ref{fig:supp:finitedist_horiz} and \ref{fig:supp:aberrations}). The experimental X-ray intensity map $I_\phi(x,z)$ transmitted through the Taylor-Couette cell is recorded by a planar sensor. This intensity depends on the absorbance of all the parts of the setup, including the PMMA cup and spindle, the suspending fluid and the particles. To isolate the particle contribution, we first acquire a reference intensity map $I_0(x,z)$ obtained with a geometry filled with the pure suspending fluid. The specific absorbance $A(x,z)$ solely due to the presence of particles throughout the geometry is then defined as:
\begin{equation}
    A (x,z) = - \log_{10} \frac{I_{\phi}(x,z)}{I_{0} (x,z)}\,.
\label{eq:calib}
\end{equation}

In a homogeneous medium, the absorbance $A$ can be directly related to the global volume fraction $\phi$ through the Beer-Lambert law:
\begin{equation}
    \label{eq:BeerLambert}
    A = (\epsilon_p - \epsilon_f)  w \phi \,, 
\end{equation}
where $\epsilon_f$ and $\epsilon_p $ are the specific extinction coefficients of the fluid and the particles, and $w$ is the thickness of the suspension slab crossed by the X-rays. In our experiments, the cross-section $w$ depends on the transverse coordinate $x$ (see Figure~\ref{fig:setup}b) and the particle volume fraction $\phi$ depends on both $r$ and $z$. The latter point implies that the absorbance $A(x,z)$ is a weighted average of $\phi$ over $r$, the radial distance to the rotation axis. A rigorous determination of the local particle volume fraction $\phi(r,z)$ requires the expression of Equation~\eqref{eq:BeerLambert} in an integral form and the application of an inverse Abel transform as detailed by Gholami~\emph{et al.}~\cite{Gholami2018}. Here, for the sake of simplicity, we shall focus on the two regions corresponding to $R_i\leq |x|\leq R_o$ that we refer to as the left and right cross-sections (CS) as depicted in Figure~\ref{fig:setup}b. We denote their local cross-sectional widths as $w(x)$. As discussed below, the particles are essentially homogeneously distributed in the horizontal plane under shear. Hence, we express $\phi(r,z)$ as the ratio $A(x,z) / [w(x) (\epsilon_p-\epsilon_f)]$, identifying $x$ with $r$ and ignoring the weighted average.

Additional parallax issues also induce a smoothing of the actual concentration maps on a typical scale $\ell \approx 8 a$ along the vertical direction. The reader is referred to Appendices~\ref{sec:app:aberrations} and \ref{sec:app:validation} for more details about the approximations used in data analysis and about calibration issues.

\begin{figure*}
    \centering
    \includegraphics[trim = {11pt 0 0 0}, clip] {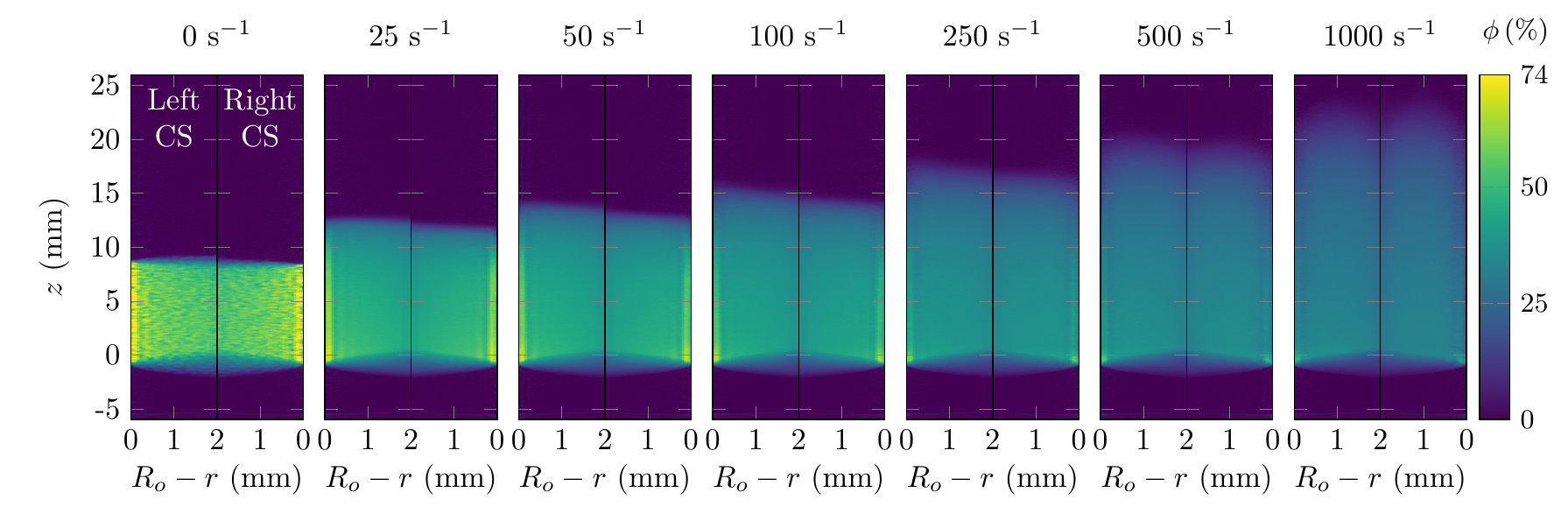}
    \caption{Local volume fraction $\phi(r,z)$ obtained by averaging X-ray images over the last 60~s of each shear rate step. The two regions of interest (left and right CS) are shown for the seven shear rates under study. Layering at the outer wall is clearly observed at rest and the lowest applied shear rates. Image blurring due to parallax issues is visible at the bottom of the geometry (see Appendix~~\ref{sec:app:aberrations}).}
    \label{fig:meanfield}
\end{figure*}

\section{Results}
\label{sec:observations}

\subsection{General Observations}

To the best of our knowledge, the concentration maps $\phi(r,z)$ shown in Figure~\ref{fig:meanfield} are the first measurements of this kind for resuspension experiments. We first notice that the particles show some verticle layering at the outer wall of the Taylor-Couette geometry. This phenomenon is particularly pronounced for the lowest shear rates where the local concentration $\phi(r,z)$ may exceed random close packing. In the particular case of the sediment at rest, it even reaches $\phi = 0.74$, indicating that particles form a hexagonal packing close to the outer wall (see leftmost panel in Figure~\ref{fig:meanfield}). We cannot conclude on the presence of layering at the inner wall: since our determination of $\phi$ close to the rotating spindle involves a weighted average on all $R_o \leq r \leq R_i$, layering may be hidden by the volume fraction in the bulk. For the same reason, we cannot completely rule out particle migration towards the outer edge: our measurements only confirm that strong migration does not occur and that the particle volume fraction is quite homogeneous in the horizontal direction except where layering is present. This result is expected in our thin-gap Couette geometry, where the stress gradients due to the curvature remain small.

In the vertical direction, starting from the top of the cup, the sediment at rest shows a rather sharp transition from $\phi = 0$ to its maximal value over the size of the typical smoothing $\ell$ resulting from the finite distance between the X-ray source and the geometry. The sediment at rest is homogeneous along $z$ with an average volume fraction over the entire gap of $0.55$, in fair agreement with the existing literature on non-attractive spheres settling very slowly~\cite{Bacri1986,Onoda1990,Dong2006,Jerkins2008}. In particular, this volume fraction is clearly lower than random close packing ($\phi = 0.636$) and than the critical volume fraction where the shear viscosity of monodisperse frictional sphere suspensions diverges (usually $0.58 \leq \phi \leq 0.62$, $0.6$ in our case)~\cite{Ovarlez2006,Boyer2011,Mari2015,Guazzelli2018}.

For increasing shear rates, Figure~\ref{fig:meanfield} shows the resuspension process in action: the top of the sediment rises as $\dot\gamma$ increases. In order to conserve the number of particles, the volume fraction in the bulk of the sediment decreases. In contrast with the sediment at rest, a finite vertical gradient in volume fraction develops throughout the sediment, including at the top of the sediment where $\phi \to 0$.

Finally, it can be noted that the concentration maps in the left and right cross-sections differ slightly from each other. Measurements of the resuspended sediment height highlight the relative differences, ranging from 1 to $9\%$ depending on the applied shear rate. These differences could stem from a slight misalignment of our geometry: our measurements show a $3\%$ --two pixels in our images-- difference in size between the two cross-sections. They did not show any significant angular misalignment between the spindle and the cup rotation axes. Any potential misalignment does not prevent steady states to be reached, as can be seen in Figure~\ref{fig:supp:spatiotemp} in Appendix~\ref{sec:app:spatiotemp}. In the following, we focus on one-dimensional profiles of the local volume fraction that we compare to predictions of the SBM based on the various expressions proposed in the literature for the normal viscosity. 

\subsection{Concentration Profiles Along the Vertical Direction}
\label{sec:extractprofiles}

\begin{figure*}
    \centering
    \includegraphics[trim = {8pt 0 0 0}, clip] {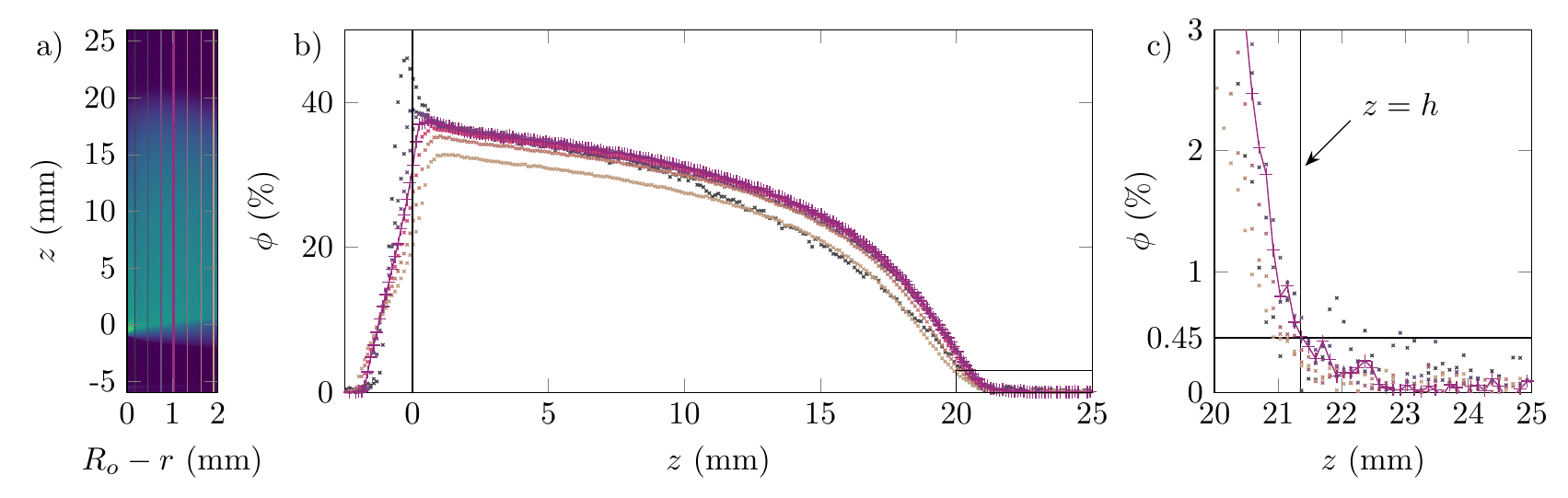}
    \caption{Computing concentration profiles along the vertical direction in the left cross-section. a) Concentration map $\phi(r,z)$ for $\dot\gamma = 500$~$s^{-1}$. Colored lines correspond to the center $r_0$ of the vertical slices over which $\phi(r,z)$ is averaged to produce one-dimensional concentration profiles $\phi(r_0,z)$. b) Concentration profiles $\phi(r_0, z)$ obtained by averaging over the seven different vertical slices, the centers of which are indicated by colored lines in Panel a), using the same color code. c) Close-up of the top region between 20 and 25~mm of the center panel. The volume fraction is considered to be above noise level when $\phi>0.45~\%$ (horizontal black line).
    }
    \label{fig:profiles_500s-1}
\end{figure*}

In order to extract one-dimensional concentration profiles from the two-dimensional maps of Figure~\ref{fig:meanfield}, we perform local averages of $\phi(r,z)$ over various vertical slices across the gap of the Taylor-Couette cell and compute $\phi(r_0,z)=\langle\phi(r,z)\rangle_{r_0-\Delta r< r \leq r_0+\Delta r}$, where $r_0$ and $\Delta r = (R_o-R_i) / 14\simeq 0.14$~mm respectively denote the centre position and the width of each slice. As shown in Figure~\ref{fig:profiles_500s-1} for the step at $\dot\gamma=500$~s$^{-1}$, the outermost and the innermost profiles significantly differ from the other profiles, confirming that layering is present at the walls and that particle migration in the bulk is limited. These profiles also confirm that, under shear, the particle volume fraction continuously increases from the clear fluid down to the bottom of the sediment rather than reaching a constant value in the bulk of the sediment.

Defining concentration profiles $\phi(r_0,z)$ requires a common reference point $z = 0$ for the bottom of the sediment. The parallax issues described in Appendix~\ref{sec:app:aberrations} imply that the volume fraction smoothly decreases to zero at the bottom of the sediment (see Figure~\ref{fig:profiles_500s-1}b) so that we cannot trivially choose $z = 0$ as the bottom location at which $\phi$ vanishes. Since they are more prominent close to the inner cylinder, we define $z = 0$ as the location of the maximum volume fraction of the second outermost volume fraction profile. We finally define the height of the resuspended sediment, named $h$, as the maximum vertical position $z$ at which the particle volume fraction exceeds the noise level of our measurements, estimated to be around $0.45\%$ (see  Figure~\ref{fig:profiles_500s-1}c). We also remark that $\phi(r_0,z)$ reaches zero with an oblique asymptote at $z=h$ i.e. $\phi(r_0,z)$ scales roughly as $(h-z)$ at the top of the sediment.

\begin{figure}
    \centering
    \includegraphics[trim = {0pt 0 0 0}, clip] {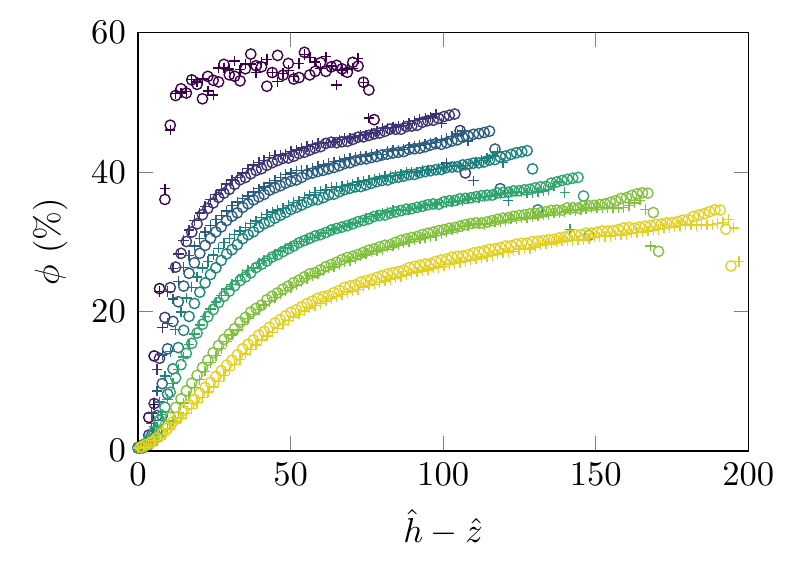}
    \caption{Concentration profiles $\phi$ derived from the central zones of each cross-section of the gap. Circles: left CS. Crosses: right CS. Colors code for the imposed shear rate from $\dot\gamma=0$~s$^{-1}$ (dark purple) to $\dot\gamma=1000$~s$^{-1}$ (yellow). Both $z$ and the resuspended sediment height $h$, as defined in Figure~\ref{fig:profiles_500s-1}c, are normalized by the particle radius $a$.
    }
    \label{fig:allprofiles_leftright}
\end{figure}

In the following, we choose to work with the central slice $\phi(r_0=24$~mm$,z)$, hereafter noted $\phi(z)$ for simplicity, as representative of the particle vertical distribution (see purple profile in Figure~\ref{fig:profiles_500s-1}). Plotting this quantity with the origin shifted to the top of the sediment shows that $\phi(z)$ follows the same trend for both cross-sections (see Figure~\ref{fig:allprofiles_leftright}). This means that particles at the top are insensitive to the particle volume fraction in the bulk of the resuspended sediment and at the bottom of the cup. This justifies the use of a local theory such as the SBM to model resuspension processes. It also means that we may focus our study on the left cross-section only without any loss of generality. 

\section{Comparison with Resuspension Models}
\label{sec:compare}

\subsection{Theoretical Framework}
\label{sec:theory}

Though initially modelled in the framework of a diffusive model~\cite{Leighton1986, Acrivos1993}, viscous resuspension may also be described using the SBM~\cite{Zarraga2000, Guazzelli2018}. The advantages of this second approach lie in that (i)~the model is fully tensorial and (ii)~the same phenomenological expressions can be used to account for particle migration, for normal forces and for particle pressure measurements. Though the latter argument is not strictly valid~\cite{Lhuillier2009,Nott2011}, it has been argued that the theoretical refinements detailed in \citet{Lhuillier2009} and \citet{Nott2011} might be neglected provided that contact forces dominate the particle stress, as should be the case at sufficiently high volume fraction~\cite{Nott2011}. We therefore choose to discuss the present experimental data in the framework of the SBM. 

The steady-state concentration profile then results from the momentum balance in the particle phase, which reads~\cite{Morris1999}: 
\begin{equation}
    \phi \Delta \rho \mathbf{g} + \mathbf{\nabla} \cdot {\Sigma_p} =0\,,
    \label{eq:sbmbase}
\end{equation}
where $\Sigma_p$ is the particle stress tensor, assumed to be diagonal~\cite{Morris1999,Guazzelli2018} and with components $\Sigma_{p,ii}=-\eta_0 \left|\dot{\gamma}\right| \eta_{n,i}\left(\phi\right) $ where $i=1$, 2 and 3. Defining the global Shields number:
\begin{equation}
    {\rm Sh} = \frac{\eta_0 \dot\gamma}{\Delta \rho g a}\,,
    \label{eq:shields}
\end{equation}
which is a constant control parameter for a given shear rate, we may recast Equation~\eqref{eq:sbmbase} in dimensionless form. Assuming that the normal viscosity coefficient $\eta_{n,3}$ is a function of $\phi$ only, we write:
\begin{equation}
   \frac{\phi}{{\rm Sh} } = -\frac{\rm d \eta_{n,3}}{\rm d \phi} \frac{{\rm d} \phi}{{\rm d} \hat{z}}\,,
    \label{eq:sbmdimensionless}
\end{equation}
where $\hat{z}=z/a$. Other lengths will be normalized in a similar fashion in the following sections.

Equation~\eqref{eq:sbmdimensionless} is formally analogous to the one obtained using the diffusive model of \citet{Acrivos1993}. By identification, we can relate the normal viscosity $\eta_{n,3}$ to the dimensionless shear-induced diffusivity $\hat{D}$ of the diffusive model, ${\rm d}\eta_{n,3} /{\rm d}\phi = 9\hat{D}/2f(\phi)$, where $f(\phi)$ is the hindrance function accounting for the presence of other particles.  

Equation~\eqref{eq:sbmdimensionless} can be solved together with an equation for mass conservation to compute the total height $\hat h$ of the resuspended sediment and the volume fraction profile, provided one assumes an expression for $\eta_{n,3}(\phi)$. Several tentative empirical expressions for this quantity --also called correlations in the literature~\cite{Acrivos1993,Guazzelli2018}-- have been proposed, which generally assume the following form:
\begin{equation}
\label{eq:normvisc_correlations}
    \eta_{n,3}(\phi) = \lambda \left ( \frac{\phi / \phi_m}{1-\phi/\phi_m} \right )^n\,,
\end{equation}
where $\phi_m$ is the volume fraction at which both shear and normal viscosities diverge. \citet{Zarraga2000} choose $n = 3$, $\lambda = 0.24$ 
and $\phi_m = 0.62$ based on previous experimental results of viscous resuspension in a Taylor-Couette geometry~\cite{Acrivos1993}. \citet{Morris1999} obtain $n = 2$, $\lambda=0.38$ and $\phi_m=0.68$ combining sets of data from large-gap Taylor-Couette and parallel-plate migration experiments. This scaling is very similar to the one proposed by \citeauthor{Boyer2011}~\cite{Boyer2011,Boyer2011a} who derived  $n=2$, $\lambda = 0.6$ and $\phi_m=0.585$ from pressure-imposed shear and rotating rod experiments.

The volume fraction profiles can also be computed using the expressions for the normal viscosity proposed by \citet{Boyer2011} and by \citet{Zarraga2000}. Indeed, Equations~\eqref{eq:sbmdimensionless} and~\eqref{eq:normvisc_correlations} allow us to derive $\hat h - \hat z$ as a function of $\phi$:
\begin{align}
    \label{eq:Boyer_z_phi}
    \hat h - \hat z &=  {\rm Sh} \frac{\lambda}{\phi_m} \frac{\phi (2\phi_m-\phi)}{(\phi_m-\phi)^2} & \text{ for \citetalias{Boyer2011},} \\
    \label{eq:Zarraga_z_phi}
    \hat h - \hat z &= {\rm Sh}  \frac{\lambda}{2\phi_m} \frac{\phi^2 (3\phi_m-\phi)}{(\phi_m - \phi)^3} & \text{ for \citetalias{Zarraga2000}}
\end{align}
Since these two expressions are strictly monotonic functions of $\phi$, we can invert them to generate numerical concentration profiles. In the case of the \citetalias{Boyer2011} correlation, inverting Equation~\eqref{eq:Boyer_z_phi} analytically results in:
\begin{equation}
    \frac{\phi(\hat z)}{\phi_m} = 1 - \left ({1 + \frac{\phi_m}{\lambda {\rm Sh}} (\hat{h} - \hat{z})} \right )^{-1/2}\,.
    \label{eq:boyerprofile}
\end{equation}

Finally we need to estimate $\hat h$ in Equations~\eqref{eq:Zarraga_z_phi} and~\eqref{eq:boyerprofile} in order to superimpose numerical profiles to experimental data. To do so, we use particle conservation,
\begin{equation}
    \hat {h}_0 \phi_{\rm m} = \int_0^{\hat h} \phi(\hat z)\,{\rm d} \hat z\,,
    \label{eq:massconservationh}
\end{equation}
together with the values of $\hat{h}_0$ obtained in Appendix~\ref{sec:app:validation} for the left and right cross-sections of the gap. While the resuspended sediment heights following the correlation of \citetalias{Zarraga2000} have to be computed numerically, we can derive a closed, explicit formula for $\hat h$ from the correlation of \citetalias{Boyer2011}:
\begin{equation}
    \hat{h}({\rm Sh}) = \hat{h}_0 + 2 \sqrt{\frac{\lambda \, \hat{h}_0}{\phi_m }{\rm Sh}}\,.
    \label{eq:Boyerheight}
\end{equation}

\subsection{Height of the sediment}

\begin{figure}
    \includegraphics{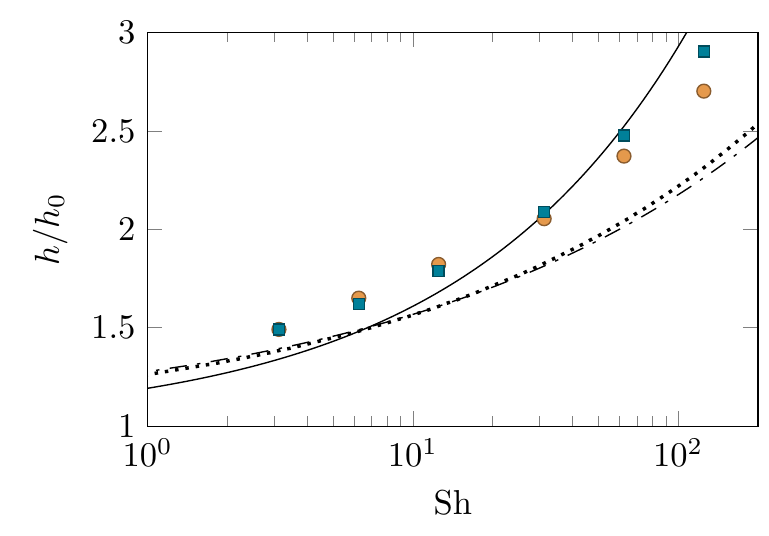}
    \caption{Variations of the relative resuspended sediment height $h/h_0$ as a function of the Shields number. Blue squares show experimental results for the left CS, and orange circles the right CS. The solid line shows the sediment height from the \citetalias{Boyer2011} correlation and given by Equation~\eqref{eq:Boyerheight}. The dotted line shows the sediment height from the \citetalias{Zarraga2000} correlation. The dash-dotted line is the sediment height predicted by \citet{Acrivos1993}.
    }
    \label{fig:resuspensionheight}
\end{figure}

We first compare our results to the resuspension heights measured by \citet{Acrivos1993}, who did not have access to local particle concentration measurements. 
We report our experimental results in Figure~\ref{fig:resuspensionheight} along with the analytical expression of Equation~\eqref{eq:Boyerheight} and two other estimates of $\hat h ({\rm Sh})$ that are very close to one another: the first is a numerical derivation based on \citeauthor{Zarraga2000} while the second is the historical asymptotic expression proposed by \citeauthor{Acrivos1993}. Our experimental data lie systematically above the \citetalias{Zarraga2000} correlation while also disagreeing with Equation~\eqref{eq:Boyerheight}. Height measurements indeed offer limited or even ambiguous insight on the processes at play and highlights the critical importance of measuring the bulk concentration profiles in the sediment in order to fully understand viscous resuspension.

\subsection{Direct estimation of the normal viscosity}
\label{sec:normalviscosity}

The volume fraction profiles measured in the previous section allow us to determine directly the normal viscosity. Indeed, integrating Equation~\eqref{eq:sbmdimensionless} from any normalized position $\hat z$ along the resuspended sediment height to $\hat h$ leads to
\begin{equation}
    \eta_{n,3}(\hat z)= \frac{1}{{\rm Sh} }\int_{\hat z}^{\hat h} \phi(u) {\rm d}u\,.
   \label{eq:normalviscosity}
\end{equation}
Interestingly, a single volume fraction profile provides an estimation of the normal viscosity on a range of $\phi$ that depends on the Shields number of the experiment. By varying the Shields number, we could not only cover a wide range of volume fractions but also test the robustness of the approach thanks to data redundancy. 

\begin{figure}
    \centering
    \includegraphics[trim = {0pt 0 0 0}, clip] {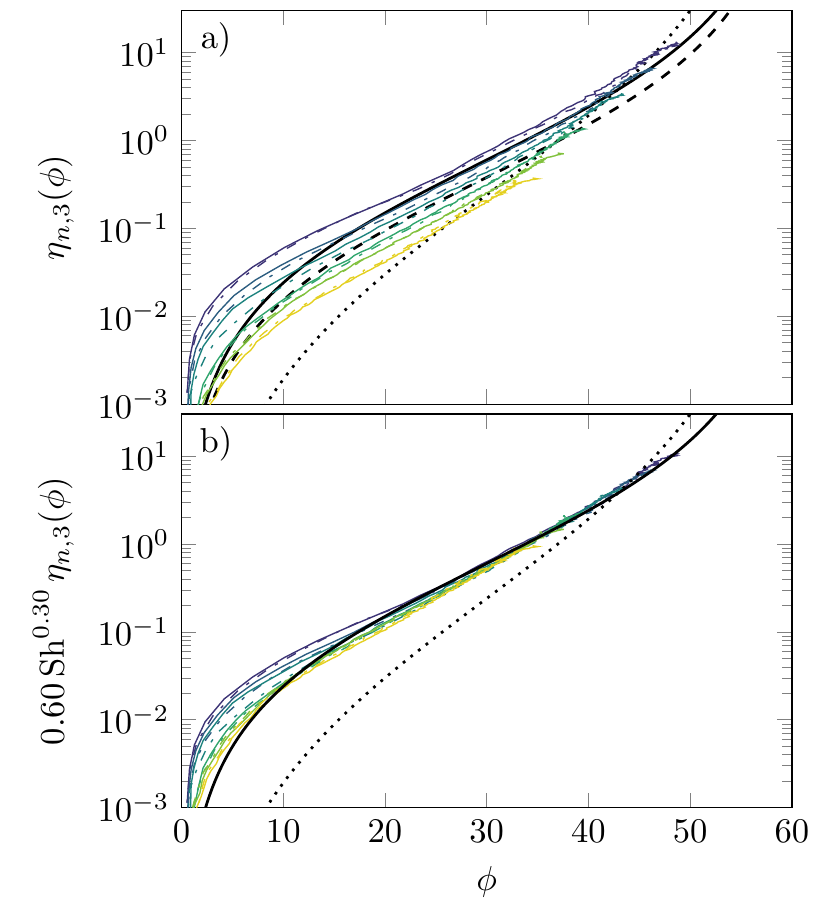}
    \caption{a) Normal viscosity $\eta_{n,3}$ as a function of volume fraction $\phi$ for shear rates ranging from $\dot\gamma = 25$~s$^{-1}$ (blue) to $\dot\gamma = 1000$~s$^{-1}$ (yellow) for the left (thin solid lines) and right (thin dash-dotted lines) cross-sections. $\eta_{n,3}$ is obtained from the experimental volume fraction profiles using Equation~\eqref{eq:normalviscosity}.
    The black solid line shows the normal viscosity correlation proposed by \citet{Boyer2011}, the dashed line corresponds to the one proposed by \citet{Morris1999} and the dotted line to the \citet{Zarraga2000} correlation. 
    b) Same data where the normal viscosity has been rescaled by a factor $0.60\, {\rm Sh}.^{0.30}$; the data coincide with the \citeauthor{Morris1999} correlation by changing the prefactor to $0.38$.
    }
    \label{fig:normalviscosities}
\end{figure}

The normal viscosities $\eta_{n,3}(\hat z)$ inferred from our experimental measurements are shown as a function of $\phi(\hat z)$ in Figure~\ref{fig:normalviscosities}a together with the correlations of \citet{Boyer2011}, \citet{Morris1999} and \citet{Zarraga2000} (see section \ref{sec:theory}). The maximum volume fraction was set to $\phi_m = 0.6$ for all correlations following the viscosity measurements reported in Appendix~\ref{sec:app:viscosity} for homogeneous suspensions. Maximum volume fractions are indeed not universal and have to be adjusted depending on the solid friction coefficient between particles~\cite{Mari2015}. Rather than collapsing on a master curve independent of the applied shear, the normal viscosity curves $\eta_{n,3}(\phi)$ shift downwards with increasing shear rates. We checked that such a shear rate dependence cannot be ascribed to the uncertainty on the local volume fraction measurements or to a lack of precision on the determination of $\hat z=0$ and $\hat z=\hat h$; it is also unlikely that potential wall slip or confinement effects both in the vertical and the lateral directions are responsible for this progressive shift (see Appendix~\ref{sec:app:slipconfinement} for more details). Moreover, none of these curves match any of the three correlations presented above in Section~\ref{sec:theory}. This result is quite intriguing since all shear rate dependence should be taken into account by the Shields number in Equations~\eqref{eq:sbmdimensionless} and~\eqref{eq:normalviscosity} for hard, spherical particles with a constant friction coefficient. 

Figure~\ref{fig:normalviscosities}b further shows that, when rescaled by a factor $0.60\, {\rm Sh}^{0.30}$, all normal viscosity profiles follow a universal shape for $\phi \geq 0.2$. Significant discrepancies are observed at low volume fractions, which can be attributed to the parallax issues described in Appendix~\ref{sec:app:aberrations}. Additional experiments shown in Appendix~\ref{sec:app:normalviscosity2g} and performed with a smaller amount of the same particles support the same scaling for $\phi \geq 30\%$. Interestingly, while the correlation proposed by \citetalias{Zarraga2000} can be clearly ruled out, the normal viscosity correlations of \citetalias{Boyer2011} and \citeauthor{Morris1999} now yield accurate descriptions of the experimental data in this new set of axes, down to volume fractions of about 0.1 for the largest shear rates, whose profiles are least sensitive to parallax issues. Such a shift amounts to using an effective Shields number in Equation~\eqref{eq:normalviscosity} given by ${\rm Sh}_{\rm eff} = 1.65\, {\rm Sh}^{0.70}$ to match the prediction of \citetalias{Boyer2011}, or $2.60\, {\rm Sh}^{0.70}$ to match that of \citeauthor{Morris1999}. This reflects the fact that the normal viscosity grows --unexpectedly-- slower than $\dot\gamma$, or, equivalently, that resuspension becomes less and less efficient as the shear rate increases. 

\subsection{Comparison with theoretical concentration profiles}
\label{sec:compar_conc}

\begin{figure*}
    \includegraphics[scale=0.97,trim = {0pt 0 0 0}, clip] {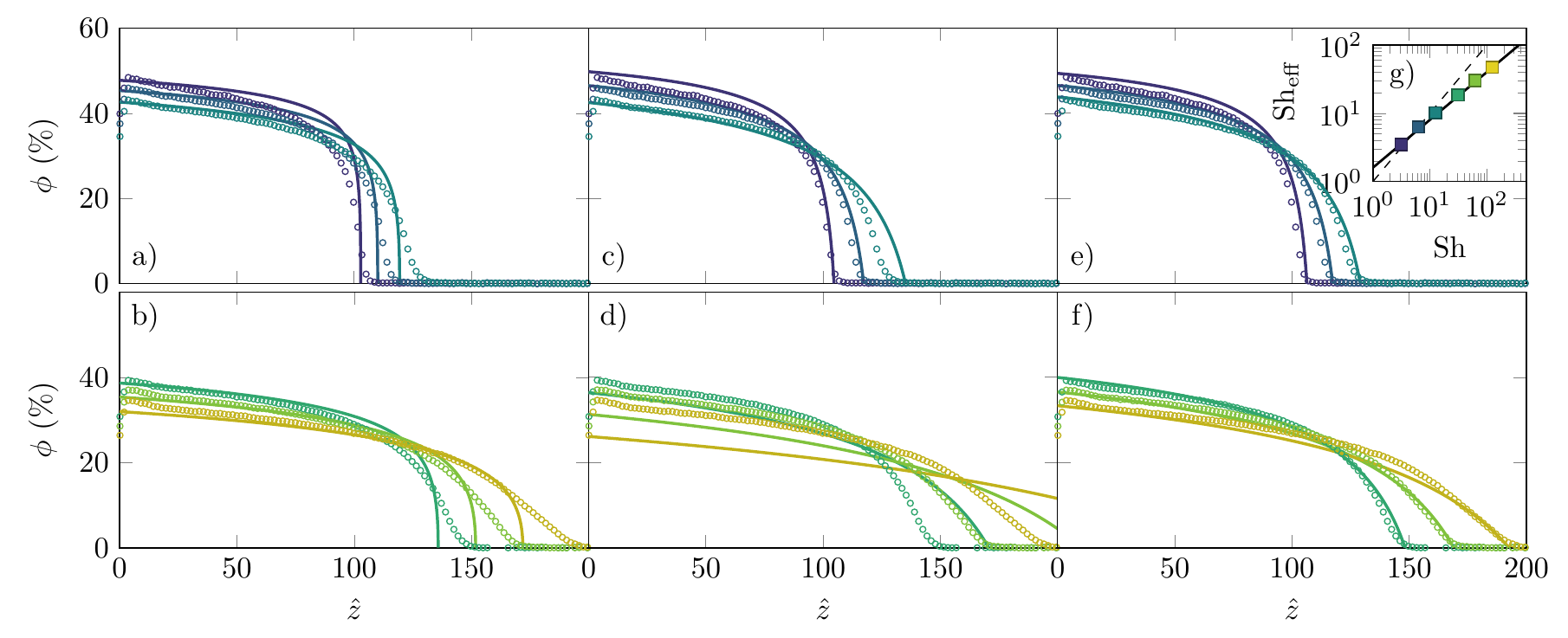}
    \caption{Particle volume fraction profiles $\phi(z)$ from the left cross-section for $\dot\gamma = 25$, $50$, $100$ [a), c) and e)] $250$, $500$ and 1000~s$^{-1}$ [b) and d) and f)]. Same color code as in Figures~\ref{fig:allprofiles_leftright} and \ref{fig:normalviscosities}. Circles correspond to experimental data whereas solid lines show the numerical profiles computed from (a-b) the \citetalias{Zarraga2000} correlation with no free parameter, (c-d) the \citetalias{Boyer2011} correlation with no free parameter and (e-f) the \citetalias{Boyer2011} correlation with the effective Shields number shown in (g) where the dashed line is ${\rm Sh} = {\rm Sh_{\rm eff}}$ and the solid line is the best power-law fit ${\rm Sh}_{\rm eff} = 1.65\,{\rm Sh}^{0.70}$.}
    \label{fig:profilefits}
\end{figure*}

Figure~\ref{fig:profilefits}a-b and \ref{fig:profilefits}c-d respectively show the numerical profiles of the volume fraction $\phi(z)$ computed from the \citetalias{Zarraga2000} and the \citetalias{Boyer2011} normal viscosity correlations without any free parameters. The \citetalias{Zarraga2000} correlation describes the sediment height $\hat h$ somewhat correctly for the three lower shear rates and underestimates it for $\dot\gamma > 100$~s$^{-1}$, as suggested in Figure~\ref{fig:resuspensionheight}. More importantly, it fails to fit the data at low volume fractions, i.e. close to $\hat z=\hat h$, where the numerical profiles display a vertical tangent whereas the experimental data show a finite slope. In contrast, the \citetalias{Boyer2011} correlation correctly fits the data at $\dot\gamma = 50$~s$^{-1}$ for almost all $z$ but grossly overestimates the sediment height at larger shear rates. 


A much better agreement is obtained if we take the Shields number as a free parameter in  Equation~\eqref{eq:boyerprofile}. The numerical profiles resulting from this procedure are displayed in Figure~\ref{fig:profilefits}e-f and coincide very well with the experimental data for all shear rates and particle volume fractions. Figure~\ref{fig:profilefits}g shows the variation of the effective Shields number deduced from the fitting procedure as a function of the global Shields number ${\rm Sh}$. Therefore, using the \citetalias{Boyer2011} correlation with an effective Shields number ${\rm Sh}_{\rm eff} = 1.65\, {\rm Sh}^{0.70}$ provides an excellent description of our data, consistently with Section~\ref{sec:normalviscosity} and, as will be discussed in the next Section, with nonlinear particle stresses as in Equation~\eqref{eq:particle_stress_thinning}.

\section{Discussion}    
\label{sec:discussion}

The volume fraction profiles obtained for a wide range of shear rates provide a strong test of the constitutive behavior of suspensions. From our observations, and after ruling out parallax issues, wall slip and confinement effects, it appears that the SBM with the standard viscous scaling of all stresses is not able to describe our system. Instead, the particle stresses seem to obey the following scaling:
\begin{equation}
\label{eq:particle_stress_thinning}
\Sigma_p \propto \left(\frac{\phi / \phi_m}{1-\phi/\phi_m} \right)^2 \dot\gamma^{\,\,0.7}\,.
\end{equation}
Let us discuss separately the dependence of the particle stress $\Sigma_p$ with respect to the volume fraction and to the shear rate. Although coupling between these two variables cannot be excluded in the general case, this separation is natural in our thin-gap Taylor-Couette geometry, as the shear rate is homogeneous. Furthermore, such an approach has been systematically used in the framework of SBM. 

\subsection{Effect of volume fraction}

As recalled in section \ref{sec:theory}, two main correlations for the normal viscosity $\eta_{n,3}(\phi)$ have been proposed in the literature. The main difference between the correlation of \citetalias{Zarraga2000} and that of \citet{Morris1999} or \citetalias{Boyer2011} is the value of the exponent $n$ in Equation~\eqref{eq:normvisc_correlations}, which sets both the asymptotic behavior at low volume fraction $\eta_{n,3} \sim \phi^n$ and the one close to jamming. The fits to the concentration profiles of Figure~\ref{fig:profilefits} and the direct normal viscosity determination of Figure~\ref{fig:normalviscosities} lead to the same conclusion: the $n=2$ correlation of \citet{Boyer2011} matches our results much better that the $n=3$ equation proposed by \citet{Zarraga2000}. In particular, close to the top of the sediment, the oblique asymptotes of the $\phi(z)$ profiles highlight the $n=2$ scaling of the normal viscosity at low $\phi$, which was proposed --yet never measured-- by \citeauthor{Morris1999} and by \citetalias{Boyer2011}.

Strikingly, the $n=3$ scaling was deduced in Ref. \onlinecite{Zarraga2000} from viscous resuspension experiments in a Taylor-Couette geometry, in which \citet{Acrivos1993} measured the resuspended sediment height. However, as shown in Figure~\ref{fig:resuspensionheight} and contrary to the volume fraction profiles, this height provides limited insight into the asymptotic behavior of the normal viscosity at low $\phi$, which might allow an $n=3$ scaling to fit the data. 


From a physical point of view, the $\phi^2$ scaling at low volume fractions points to pairwise interactions, contrary to a $\phi^3$ scaling which would rather hint at three-body hydrodynamic interactions as considered in Ref. \onlinecite{Wang1998}. As pointed out by Lhuillier \cite{Lhuillier2009}, both non-hydrodynamic and hydrodynamic forces could in principle lead to shear-induced migration. It has been argued that the effect of non-hydrodynamic forces such as contact forces between particles is likely to dominate at high volume fraction \cite{Nott2011}. Our finding that $\Sigma_p\propto\phi^2$ at low $\phi$ thus indicates that the shear-induced migration evidenced in resuspension experiments originates --at least for the system studied here-- from non-hydrodynamic interactions, even at low volume fractions, where the shear viscosity is dominated by viscous stress. As will be shown in the following, these non-hydrodynamic interactions are likely to be contact forces. 


\subsection{Effect of shear-rate: nonlinear particle stress}

The nonlinear power-law scaling of the particles stresses $\Sigma_p\propto\dot\gamma^{0.7}$ [Equation~\eqref{eq:particle_stress_thinning}] is not expected for a viscous suspension when interparticle contact forces are modeled by an ideal Coulomb friction law~\cite{Guazzelli2012}. Many experimental results, however, show departures from the simple viscous scaling $\Sigma_p\propto\dot\gamma$ through a shear-thinning viscosity at high volume fraction in non-Brownian suspensions~\cite{Zarraga2000,Dbouk2013,Chatte2018,Tanner2018,Lobry2019}. For our suspension, we find in Appendix~\ref{sec:app:viscosity} that $\eta\propto\dot\gamma^{-0.17}$ for the largest volume fractions. Recently, several authors have introduced non-Coulomb friction laws to explain such shear thinning, either by adding an explicit velocity dependence~\cite{Tanner2018} or a normal stress dependence~\cite{Chatte2018,Lobry2019} to the microscopic sliding friction coefficient between particles. 

Still, the shear viscosity cannot be correlated directly to the particle stress that we extract from the shear-induced resuspension experiment. Indeed, the shear viscosity of the suspension finds its contribution from both hydrodynamic and contact stresses. In order to get some insight into contact stresses, one needs to perform, for example, shear reversal experiments~\cite{Gadala-Maria1980,blanc2011local,Lin2015,Peters2016}. In such experiments, the suspension is first sheared at a given shear stress or shear rate in a simple shear flow until steady-state is reached. The direction of shear is subsequently reversed (keeping the same value of the applied shear rate or stress), and the viscosity evolution with strain is recorded. In the presence of contact forces, most contacts are predominantly oriented in the compression direction of simple shear. Upon shear reversal, all of these contacts are now in traction and are suddenly broken, which results in an abrupt decrease of the viscosity. The minimum viscosity achieved during the reversal can then be associated mostly with hydrodynamic interactions, whereas the difference between the steady-state viscosity and this minimum provides the contact contribution to the viscosity~\cite{Peters2016}. 

In order to estimate the particle contact contribution to the shear viscosity, we performed shear reversal experiments on suspensions composed of the same glass beads as in the resuspension experiments, but in a more viscous fluid to avoid sedimentation. These experiments are discussed in Appendix~\ref{sec:app:reversal} and the results are shown in Figure~\ref{fig:supp:viscosity_reversals}. Following \citet{Peters2016} to analyze these results, we show that the hydrodynamic part of the shear viscosity is essentially rate-independent, whereas the contact contribution to the viscosity shows a pronounced shear-thinning behavior (see Table~\ref{tab:supp_thinning_exponents}). It suggests that the contact contribution to stress scales as $\dot\gamma^{0.76}$, which is broadly consistent with particle stresses scaling as $\dot\gamma^{0.7}$ in resuspension experiments [Equation~\eqref{eq:particle_stress_thinning}]. Altogether, these observations point to non-Coulomb friction between our glass particles, as recently proposed by \citet{Chatte2018} and by \citet{Lobry2019}. This means that the resuspension properties of non-Brownian particles should depend much on the exact nature of the particles and on the way their friction coefficient varies with load and velocity.

\section{Conclusion}

Previous experimental studies of shear-induced resuspension in the literature focused on the height of the resuspended sediment and needed to assume a viscous scaling either for the particle stress (in the case of the SBM) or for the diffusion coefficient (in the case of the Acrivos model) in order to model their observations. This assumption is justified in the case of rate-independent Coulomb friction between the particles. Here, we have obtained local volume fraction profiles for resuspension in a Taylor-Couette geometry thanks to X-ray imaging. A broad range of volume fractions is covered under various applied shear rates, which allows us to investigate both the volume fraction and shear rate dependence of particle stresses. In the framework of the SBM, our data demonstrate that the particle stresses asymptotically scale as $\phi^2$ at low volume fractions and display a nonlinear, shear-thinning scaling with respect to the shear rate. The latter is consistent with the shear thinning observed both in the shear viscosity of the suspension and in the contribution of contacts to this viscosity. This likely points to a non-Coulomb or velocity-weakening friction between the glass particles. Similar experiments should now be conducted on other non-Brownian particles in order to get more insight into the impact of the precise local contact laws between pairs of particles on the particle stresses, as well as on the efficiency of resuspension. 

\begin{acknowledgments}
This research was supported in part by the Agence Nationale de la Recherche through the FLUIDIDENSE project (grant number ANR-17-CE07-0040). We wish to acknowledge the support of the Plateforme de Caractérisation des Matériaux d'Aquitaine (PLACAMAT) for allowing experimental time on the X-ray tomography equipment. We thank Nicolas Lenoir and Ronan Ledevin for technical support during the experiments. We also thank \'Elisabeth Guazzelli, Elisabeth Lemaire and Fr\'ed\'eric Blanc for insightful discussions. 
\end{acknowledgments}

\clearpage
\appendix

\section{Macroscopic behavior}
In this Appendix we report and discuss the steady-state macroscopic viscosity of suspensions. We also present the results of a series of shear reversal experiments performed to characterize the particle contact contribution to the shear viscosity. 

We use the same glass spheres as those used in the resuspension experiments of the main text. This time, the particles are suspended in a mixture of 20\% wt. water and 80\% wt. UCON (instead of 65\% wt. water, 35\% wt. UCON), which is also a Newtonian fluid, with a viscosity $\eta_0\simeq 14.5$~Pa.s$^{-1}$ at 25$^\circ$C. Suspensions are prepared at various volume fractions ranging from 10 to 50\%. 
We use a Kinexus Ultra+ rheometer (Malvern Panalytical) equipped with a wide-gap Taylor-Couette geometry, with sandblasted surfaces (rotor diameter 25~mm, stator diameter 37~mm), to characterize the suspensions. Experiments are conducted in the controlled-stress mode. The shear stress and shear strain are obtained from the applied torque and the measured rotation angle by using the standard equations for the Taylor-Couette geometry at the rotor surface.

\subsection{Steady-shear viscosity}\label{sec:app:viscosity}
For each volume fraction, the steady-shear apparent viscosity is measured by applying a series of logarithmically-spaced constant shear stresses, corresponding to shear rates varying between 0.01 and 10 s$^{-1}$. Each shear stress is applied for a duration of 10~s. 

\begin{figure}[h]
    \centering
    \includegraphics{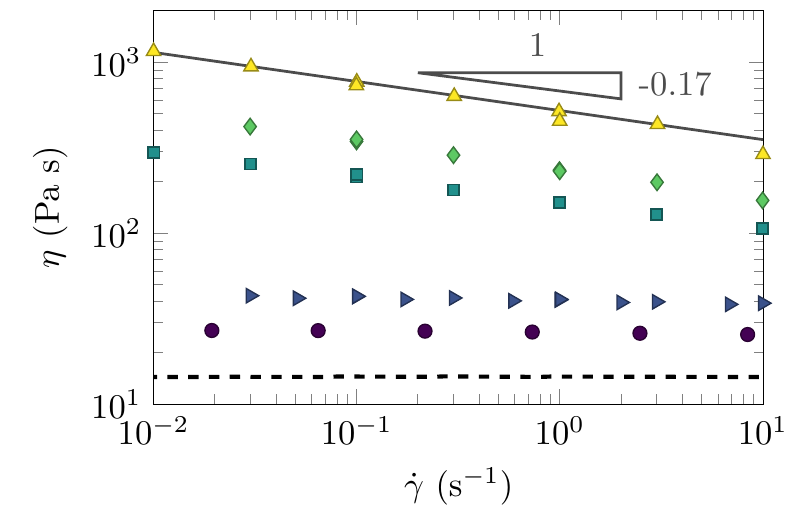}
    \caption{Shear viscosity $\eta$ of a suspensions of glass spheres (diameter $2a = 250~\mu$m) in an 80\% wt. UCON aqueous solution as a function of the shear rate $\dot\gamma$. The thick black line shows the viscosity of the pure fluid. The purple circles, blue left pointing triangles, teal squares, green diamonds and yellow upwards pointing triangle respectively correspond to $\phi = 10 \%$, 20\%, 40\%, 45\% and 50\%. The gray solid line provides the best power-law fit $\eta \propto \dot\gamma^{-0.17}$ to the $50\%$ data.}
    \label{fig:supp:viscosity_vs_rate}
\end{figure}

The steady-state viscosity $\eta$ is plotted as a function of the steady-state shear rate in Fig.~\ref{fig:supp:viscosity_vs_rate} for various particle volume fractions $\phi$. At low $\phi$, a Newtonian behavior is observed with a viscosity higher than that of the interstitial fluid. As the volume fraction is increased, the viscosity increases and a mild shear-thinning is observed. For $\phi=50\%$, the apparent viscosity follows a scaling $\eta\propto\dot\gamma^{-0.17}$. Similar shear-thinning at high particle concentration has already been reported in a number of other non-Brownian suspensions~\cite{Zarraga2000,Dbouk2013,Chatte2018,Tanner2018, Lobry2019}. 

\begin{figure}[h]
    \centering
     \includegraphics {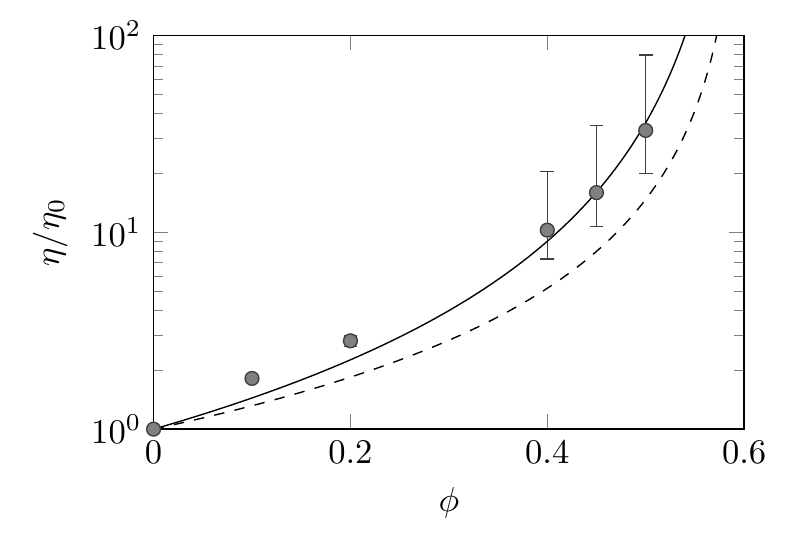}
    \caption{Dimensionless shear viscosity $\eta / \eta_0$ of a suspension of glass spheres (diameter $2a=250 \mu$m) in an 80\% wt. UCON aqueous solution, plotted for $\dot\gamma = 1$~s$^{-1}$, as a function of the particle volume fraction $\phi$. Error bars on the experimental data show the maximum and minimum value measured over the range of shear rates of Fig.~\ref{fig:supp:viscosity_vs_rate}. The empirical viscosity laws of Maron and Pierce~\cite{Maron1956} (solid line) and Krieger and Dougherty~\cite{Krieger1959} (dashed line), both with $\phi_m=0.6$, are also shown.}
    \label{fig:supp:viscosity_vs_phi}
\end{figure}

In order to characterize the viscosity increase with volume fraction, we plot the dimensionless viscosity $\eta(\phi, \dot\gamma = 1~{\rm s}^{-1})/\eta_0$ as a function of $\phi$ in Figure~\ref{fig:supp:viscosity_vs_phi}. Since the suspensions show shear-thinning at high concentrations, we also show error bars extending from the lowest to the highest dimensionless viscosity measured at each $\phi$ within the range of investigated shear rates. The increase of $\eta / \eta_0$ with $\phi$ is similar to that reported for other non-Brownian suspensions~\cite{Guazzelli2018} and may be fitted fairly well to a Maron-Pierce law~\cite{Maron1956}:
\begin{equation}
    \frac{\eta}{\eta_0} =\left (1-\frac{\phi}{\phi_m} \right)^{-2}\,,
\end{equation}
with $\phi_m=0.6$, which is in the range of $\phi_m$ values typically observed for monodisperse frictional spheres~\cite{Ovarlez2006,Boyer2011,Mari2015,Guazzelli2018}.

\begin{figure*}
    \centering
    \includegraphics[scale=0.80] {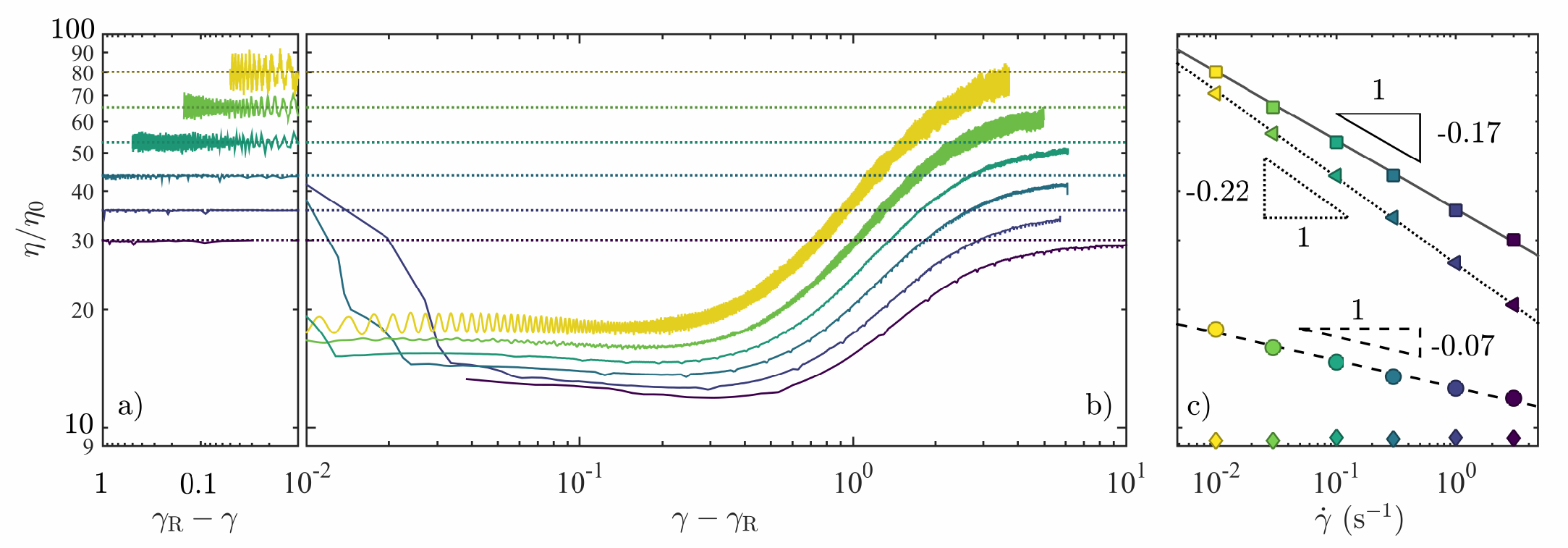}
    \caption{Shear Reversal experiments of a 50 \% wt. suspension of glass particles in an 80 \% wt. UCON-20\% wt. aqueous mixture. a) Steady-state relative viscosity $\eta/\eta_0$ of the suspension plotted as a function of the strain before reversal $\gamma_{\rm R} - \gamma$. Colors code for the applied shear stress ranging from 11.5~Pa (yellow) to 1310~Pa (purple). b) Evolution of the relative viscosity $\eta/\eta_0$ as a function of the shear strain after reversal $\gamma - \gamma_{\rm R}$. The shear reversal data for the all shear rates were smoothed over 10 samples for clarity. c) Relative viscosity $\eta/\eta_0$ as a function of the applied shear rate. Squares: steady-state viscosity, obtained from a). Circles: relative viscosity minimum $\eta_{\rm min}/\eta_0$ reached in b) soon after reversal. Triangles: contact contribution to the suspension viscosity $\eta^{\rm C}$. Diamonds: hydrodynamic contribution to the suspension viscosity $\eta^{\rm H}$. Contact and hydrodynamic contributions are deduced from the decomposition proposed by~\citet{Peters2016}, which we recall in Equations~\eqref{eq:supp:peters_decompose1} and~\eqref{eq:supp:peters_decompose2}. Same color codes as in a) and b).}
    \label{fig:supp:viscosity_reversals}
\end{figure*}

\subsection{Shear reversal experiments}
\label{sec:app:reversal}
Next, following \citet{Lin2015} and \citet{Peters2016}, we use shear reversal experiments to evaluate the particle contact contribution to the shear viscosity. As in \citet{blanc2011local}, since we use a stress-controlled rheometer, we work in the controlled-stress mode in order to monitor accurately the transient evolution of the viscosity with strain. We first shear the suspension at a constant imposed stress $\tau$ until the shear strain $\gamma$ exceeds 10 and a steady state is reached. We then apply a resting period of 10~s at zero stress before applying a stress $-\tau$ until a new steady state is reached. We call $\gamma_{\rm R}$ the reversal shear strain attained when the stress is set to $-\tau$. We monitor the evolution of the shear viscosity both before and after shear reversal. These measurements are repeated for the same values of $\tau$ as in the previous section and shown in Fig.~\ref{fig:supp:viscosity_reversals}a-b for a suspension of volume fraction $\phi=50\%$.

As already observed in the literature~\cite{Gadala-Maria1980,Lin2015,Peters2016}, the shear viscosity just after reversal is smaller than the steady-state viscosity and it subsequently increases with strain until it reaches a steady state for a strain of order 5. Figure~\ref{fig:supp:viscosity_reversals}b-c shows that the steady-state viscosity is more sensitive to the shear rate than the viscosity minimum $\eta_{\rm min}$ reached shortly after shear reversal. This last observation is consistent with the numerical simulations of \citet{Peters2016}, which suggest the following decomposition: 
\begin{align}
        \label{eq:supp:peters_decompose1}
    \eta &= \eta^{\rm H} + \eta^{\rm C}\, , \\
        \label{eq:supp:peters_decompose2}
    \eta_{\rm min} &= \eta^{\rm H} + 0.17 \eta^{\rm C} \, ,
\end{align}
$\eta^{\rm H}$ and $\eta^{\rm C}$ being the hydrodynamic and contact contributions to viscosity; the choice of the 0.17 numerical factor may \emph{a priori} depend on the interparticle, local friction coefficient. We perform a similar decomposition in Figure~\ref{fig:supp:viscosity_reversals}c. Using a numerical factor 0.12 (instead of 0.17) in Equation~\eqref{eq:supp:peters_decompose2}, we observe that the hydrodynamic part of the suspension viscosity $\eta^{\rm H}$ is constant, as expected of a suspension of non-Brownian hard spheres. Following that choice, the contact viscosity $\eta^{\rm C}$ decreases for increasing shear rates $\dot\gamma$, and point to an average shear-thinning exponent of $-0.24 \pm 0.02$ for the contact viscosity, as shown in Table~\ref{tab:supp_thinning_exponents}. This exponent broadly agrees with the $-0.30$ shear-thinning exponent of the normal viscosity obtained from resuspension experiments in the main text.

\begin{table}[h]
    \centering
    \renewcommand{\arraystretch}{1.3}
    $\begin{array}{l rrrrr}
        \hline
         \phi                &  0.10     & 0.20     & 0.40  & 0.45  & 0.50    \\
         n~({\rm total})     &  -0.01    & -0.02    & -0.15 & -0.18 & -0.17   \\
         n~({\rm contacts})  &  /~~~     &   /~~~   & -0.25 & -0.25 & -0.22   \\
         \hline
    \end{array}$
    \caption{Shear-thinning exponents $n$ of the suspension viscosity $\eta$ with the shear rate: $\eta \propto \dot\gamma^n$. The total exponent is obtained from the flow curves of Fig.~\ref{fig:supp:viscosity_vs_rate}. The contact exponent is deduced from the viscosity decomposition of \citet{Peters2016} for shear reversal experiments, such as the one shown in Fig.~\ref{fig:supp:viscosity_reversals}. A viscous scaling implies $n = 0$.}
    \label{tab:supp_thinning_exponents}
\end{table}

\section{Effect of slip and confinement in the experiment}
\label{sec:app:slipconfinement}

The results shown in Figures~\ref{fig:normalviscosities} and~\ref{fig:profilefits}, discussed in Section~\ref{sec:discussion}, are striking and counter intuitive. We must ensure that they do not result from undesirable physical effects present in our experiment. 

\subsection{Wall slip}
Slip may be present at the walls of our Taylor-Couette cell. Experiments by \citet{Jana1995} conducted in suspensions of non-Brownian hard spheres in a Newtonian solvent show that slip becomes noticeable only for volume fractions above $0.45$. The slip rate then increases with $\phi$ but remains quite limited up to $\phi = 0.52$, the maximum volume fraction investigated by \citet{Jana1995}. In our experiments, the profiles of $\eta_{n,3} (\phi)$ deviate much more strongly from the Boyer correlation at high shear rates, where the volume fraction $\phi$ is lower than at low shear rates. Therefore, it is very unlikely that such a deviation results from wall slip.

\subsection{Confinement}
The peculiar dependence of $\Sigma_p$ with $\phi$ and $\dot\gamma$ could also be due to particle confinement present in our geometry. Confinement effects may come from the narrow gap width of the geometry or from the limited height of the sediment $h$ compared to the particle radius $a$. 

\paragraph{Lateral confinement} Our experimental configuration involves $\sim 8$ particles across the gap, which is slightly below the classical limit requiring at least $\sim 10$ particle diameters to accurately reflect bulk behavior in granular materials~\cite{Andreotti2013}. However, radial confinement may not be that crucial since we report layering that extends over only 1 to 2 particle radii from the walls without any other significant radial gradients in the volume fraction. 

\paragraph{Vertical confinement} Turning to the case of vertical confinement, we note that imposing a zero velocity at the bottom of the cup provides incompatible boundary conditions with the velocity of the inner cylinder that imposes the global shear rate. The consequences of such incompatibility should be even more drastic for smaller sediment heights. To test this idea, we performed additional resuspension experiments with a number of particles two times smaller than in previous measurements (see Appendix~\ref{sec:app:normalviscosity2g}). Such experiments show a larger dispersion of the normal viscosity $\eta_{n,3}$ when plotted as a function of $\phi$. Yet, the scaling proposed in Equation~\eqref{eq:particle_stress_thinning} remains valid for the larger volume fractions, i.e. close to the bottom wall, suggesting that it results from a bulk property of the suspension rather than from confinement effects. A better control of boundary conditions at the edges of the geometry (top or bottom) is possible, for instance by using a non-miscible, very dense fluid such as mercury at the bottom of the geometry~\cite{Acrivos1993,DAmbrosio2019} or by considering positively buoyant particles in a dense fluid. Such experiments could confirm the general nature of Equation~\eqref{eq:particle_stress_thinning}.

\section{Geometrical Approximations and Parallax Issues}
\label{sec:app:aberrations}

In this Appendix, we provide more details on the approximations made in Section~\ref{sec:Xradio}. We first describe how computing $\phi(r,z)$ involves averaging over different radial positions. We then detail the impact of a finite distance between the X-ray source, the geometry and the detector on the final measurements, which we shall refer to as parallax issues in the main text. 

\subsection{General Approximation}

For an X-ray source located at infinity, the apparent thickness $w(x)$ of the slab of suspension crossed by the X-rays at a position $x$ within the two cross-sections defined in Figure~\ref{fig:setup}b is simply given by:
\begin{equation}
    w(x) = 2 \sqrt{R_o^2 - x^2}\,.
\end{equation}
X-rays actually cross slabs of the suspension corresponding to multiple values of the radial distance $r$. Given the notations in Figure~\ref{fig:setup}b, the integral version of the Beer-Lambert law reads:
\begin{equation}
\label{eq:zetaofr}
    A (x,z) = 2 (\epsilon_p - \epsilon_f) \int_{r = x}^{R_o} \phi(r,z) \underbrace{\frac{r}{\sqrt{r^2-x^2}}}_{\zeta(x,r)} {\rm d}r \,.
\end{equation}
The approximation that we make in the main text consists in assuming that: 
\begin{equation}
    \label{eq:approximation}
     \phi(r=x,z) \simeq \frac{A(x,z)} {w(x) (\epsilon_p - \epsilon_f)} \,.
\end{equation}
Since $2\int_x^{R_o}\zeta(x,r){\rm d}r=w(x)$, Equation~\eqref{eq:zetaofr} shows that the above approximation is true only for $r$-independent concentration fields or for $x$ very close to the outer edge. In the latter case, however, the precision on the volume fraction measurement is poor due to the small value of $w(x)$ (see Appendix~\ref{sec:app:validation} for more details).

Consequently, the apparent $\phi(r,z)$ inferred from Equation~\eqref{eq:approximation} is actually a weighted average of the true $\phi(r,z)$ over $x\leq r \leq R_o$. The weight of $r \simeq x$ in this averaging process is particularly high since $\zeta(x, r)$ diverges for $r \to x$. This means that in the absence of strong radial gradients in the particle volume fraction, we may use Equation~\eqref{eq:approximation} to obtain a reasonable estimate of the local volume fraction. In any case, our study is mostly focused on the vertical distribution of particles so that the details of the volume fraction field in horizontal planes can be averaged out.

\subsection{Finite distance effects: parallax issues}

In practice, both the X-ray source and the detector are located at a finite distance from the Taylor-Couette geometry. This implies that the incident X-rays, depicted so far as a parallel beam, actually diverge from the source. This has a direct impact on the data especially in the vertical plane.

\subsubsection{Changes in the horizontal plane}

\begin{figure}
    \centering
    \includegraphics[scale=0.80]{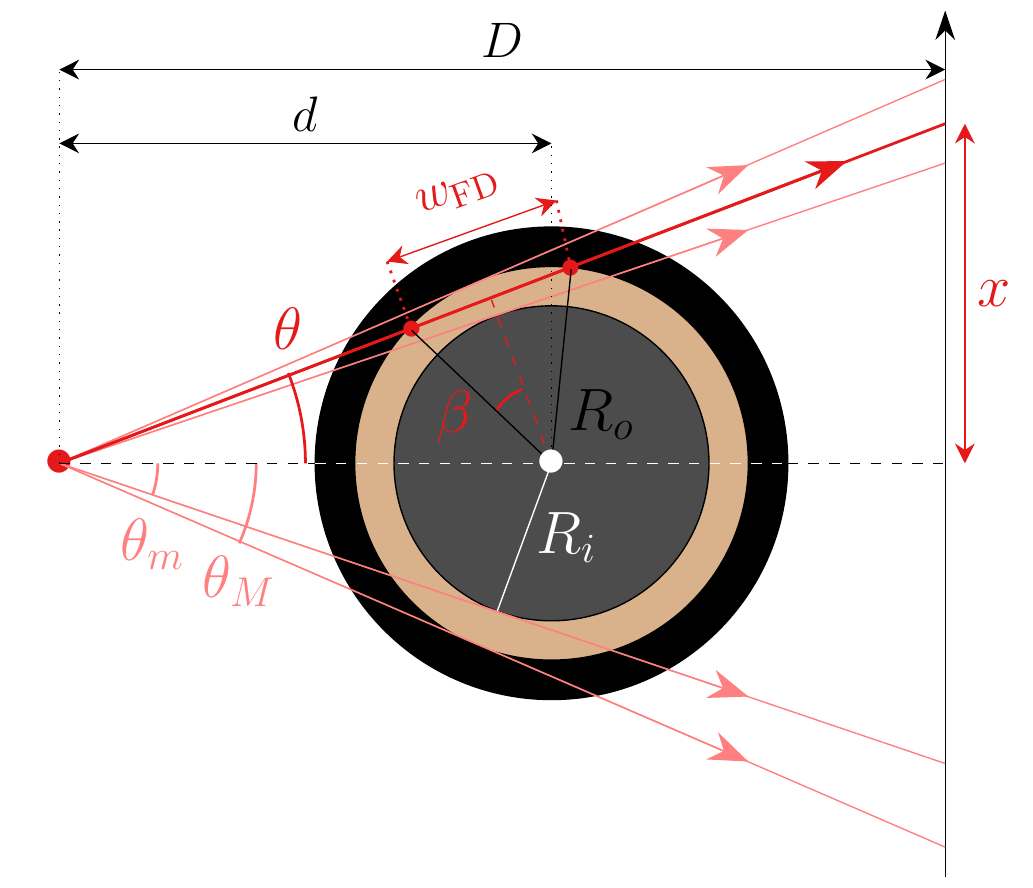}
    \caption{Impact of the finite distance between source, geometry and detector in the horizontal plane. We notice that $x$ is proportional to $D$}
    \label{fig:supp:finitedist_horiz}
\end{figure}

First, the width $w(x)$ has to be computed for a source located at a finite distance $d \simeq 25$~cm from the center of the geometry. The location $x$ at which an X-ray hits the sensor now varies with the distance between the source and the detector, $D \simeq 70$~cm. The geometrical relations shown in Figure~\ref{fig:supp:finitedist_horiz} allow us to derive the finite-distance slab thickness of the sediment crossed by the X-rays as a function of $x$ through the use of $\theta$, defined as $\tan\theta = x/D$:
\begin{align}
    w_{\rm FD}(x) &= 2 R_o \sin\beta \\
                  \label{eq:wfd}
                  &= 2  \sqrt{R_o^2 - \frac{d^2 x^2}{D^2 + x^2}} \,.
\end{align}
The boundaries of our cross-sections are defined by two limit rays of angles $\theta_m$ and $\theta_M$, such that $d \sin \theta_m = R_i$ and $d \sin \theta_M = R_o$. We can relate these angles to the locations $x_m$ and $x_M$ where the limit rays hit the sensor plane:
\begin{align}
    \label{eq:gapwidth_finitedistance}
    x_m &= \frac{R_i D}{\sqrt{d^2-R_i^2}} \,, \\
    x_M &= \frac{R_o D}{\sqrt{d^2-R_o^2}} \,.
\end{align}
We notice that $x_M - x_m$ is not equal to $R_o-R_i$ and is instead proportional to $D$. The size --expressed as a number of pixels-- of the cross-section in our raw images thus depends on both the true resolution of the sensor (the number of pixels per mm) and the distance $D$. In practice, we choose a scaling factor $S$ in the pictures so that the apparent cross-section size is equal to $S (x_m - x_M) = R_o - R_i$. 

Finally, the impact of the finite distance between the source and the detector can be quantified as $[w_{\rm FD}(S x) - w(x)] / w(x)$. Given our estimates for $d / R_o \geq 10$ and $D/R_o \geq 20$, the finite distance leads to corrections on $w$ that are always smaller than 0.1\% and can therefore be neglected.

\subsubsection{Impact in the vertical direction}
\begin{figure}[h]
    \centering
    \includegraphics[scale=0.75] {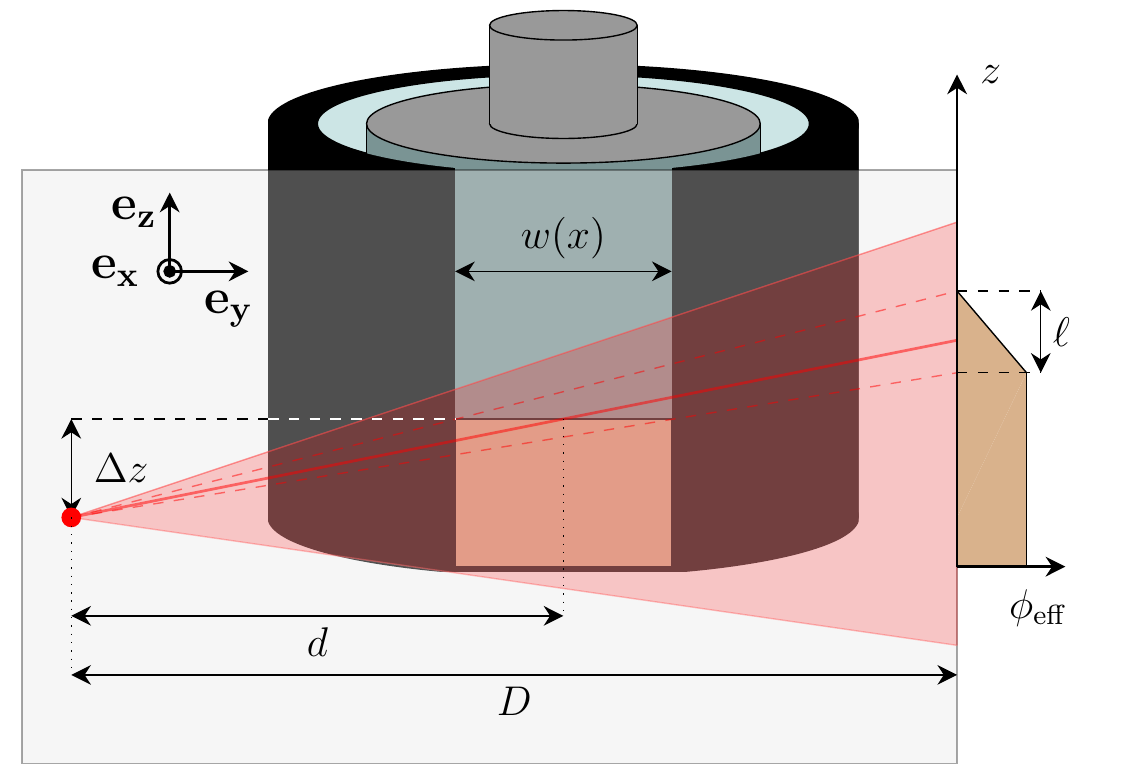}
    \caption{Parallax issues along the vertical direction. For a pointlike source located at a vertical distance $\Delta z$ from the suspension slab under study, volume fraction measurements are smoothed over a typical size $\ell$, proportional to the suspension thickness crossed by the X-rays $w(x)$ and to $\Delta z$ for small X-ray angles.}
    \label{fig:supp:aberrations}
\end{figure}

Since the source is pointlike and set at a finite distance from the geometry, X-rays may cross the suspension at some angle relative to the horizontal direction. Consequently, as shown in Figure~\ref{fig:supp:aberrations}, our measurements are also averaged vertically over a typical ``smoothing'' length $\ell$ that can be expressed as a function of the vertical distance $\Delta z$ between the source and the slab of suspension under investigation and as a function of the slab thickness $w(x)$:
\begin{equation}
    \ell =  4 D \frac{ w(x) |\Delta z|}{4 d^2 - w^2(x)}\,.
\end{equation}
In our setup, the vertical position of the source is $z \simeq 16$~mm in the reference frame of Figure~\ref{fig:meanfield}. In the case of the sediment at rest, there should be a sharp transition from $\phi = \phi_m$ to $\phi = 0$ at the top of the sediment. We can thus readily estimate $\ell$ by measuring the typical width of the transition zone on the experimental map of Figure~\ref{fig:meanfield}a. In the worst-case scenario, i.e. at the bottom of the sediment and close to the inner wall, we predict $\ell \simeq 3.3$~mm, which is compatible with the concentration maps of Figure~\ref{fig:meanfield}. We further predict $\ell \simeq 1.4$~mm at the top of the sediment at rest and close to the inner wall, in fair agreement with Figure~\ref{fig:meanfield}a-b. We finally compute an estimate of $\ell$ at the centre of the cross-section, used to derive the concentration profiles examined in Sections~\ref{sec:extractprofiles} and~\ref{sec:compare}. We obtain $\ell = 0.146 | \Delta z |$ which amounts to 4 particle diameters at the top of the sediment and to 9 particle diameters at the bottom of the sediment. The impact of parallax is very limited at the bottom of the sediment since no significant variations of $\phi$ are observed over $\ell$. Rather, its impact is maximal at the top of the sediment and for low shear rates where $\phi$ show the strongest spatial variations. Therefore, parallax issues along the vertical direction may explain the dispersion of the normal viscosity data at low volume fraction in Figures~\ref{fig:normalviscosities} and~\ref{fig:supp:normalviscosity_2g} as well as the slight curvature of the concentration profiles for $\phi$ very close to 0 in Figures~\ref{fig:profiles_500s-1}, \ref{fig:allprofiles_leftright} and \ref{fig:profilefits}. We believe the impact of this parallax issue remains limited to such details.   

\section{Data validation and calibration}
\label{sec:app:validation}

\subsection{Locating the gap in X-ray images}

When computing $\phi(r,z)$ using Equation~\eqref{eq:approximation}, we notice that both $A$ and $w$ tend to zero for $x = R_o$. Therefore, volume fraction measurements close to the outer cylinder are particularly sensitive to where we define $x = R_o$ in the raw images. Any imprecision $\delta x$ on the position of the outer wall may result in dividing $A$ by an effective thickness $w(x+\delta x)$ that greatly differs from the actual $w(x)$ in the vicinity of $x = R_o$. Thus, we carefully choose the position of $x=R_o$ in both cross-sections so as to (i)~obtain a gap width compatible with the actual gap of the geometry (we respectively get $2.00$~mm and $2.06$~mm for the left and right cross-sections) and (ii) obtain a local minimum of volume fraction at the outer edge of the geometry, followed by a progressive increase of $\phi$ up to $r \approx a$ for all shear rates. Our setup indeed allows us to resolve particles that are in contact with the walls, which naturally introduces an apparent particle concentration gradient at the outer boundary.

\subsection{Estimating the extinction coefficients $\epsilon_p$ and $\epsilon_f$}

\begin{figure}
    \centering
    \includegraphics[scale=1] {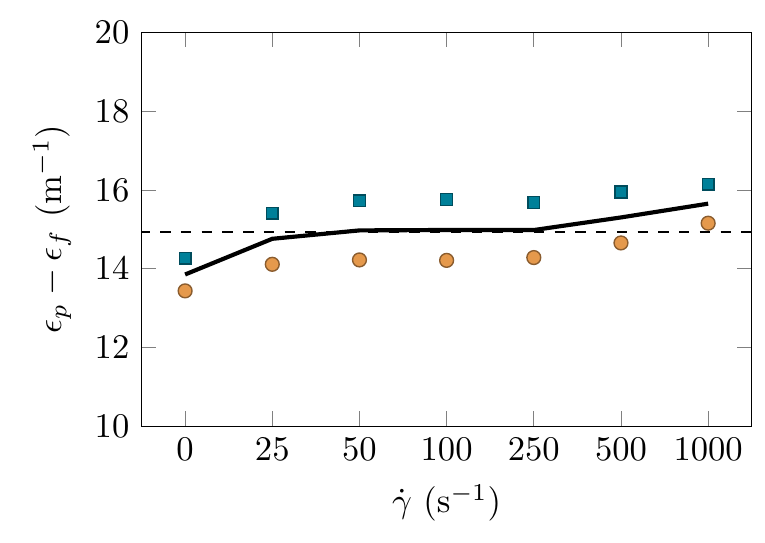}
    \caption{Estimation of the differential extinction coefficient $\epsilon_p - \epsilon_f$ deduced from the concentration profiles when using Equation~\eqref{eq:A_to_phi} with the total particle mass $M = 4.00$~g. Blue squares: estimation from the left CS. Beige circles: estimation from the right CS. The black line shows the average over the two CS. The dashed line corresponds to the global average $\epsilon_p - \epsilon_f = 14.9$~m$^{-1}$.
    \label{fig:supp:estimatedepsilon}
    }
\end{figure}

The last step when converting $A(x,z)$ into $\phi(r,z)$ consists in estimating the proportionality constant between $A/w$ and $\phi$, namely the extinction coefficient difference $\epsilon_p - \epsilon_f$. To this aim, we perform the following integral:
\begin{equation}
    2 \pi \rho_p \int_{R_i}^{R_o} x {\rm d}x \int_{0}^{\infty}  {\rm d}z \frac{A(x,z)}{w(x)} = M (\epsilon_p - \epsilon_f)
    \label{eq:A_to_phi}
\end{equation}
For a given shear rate, we compute the above integrals for each cross-section. Measuring independently the particle mass $M=4.00 \pm 0.01$~g with a precision scale allows us to deduce the extinction coefficient difference $\epsilon_p-\epsilon_f$ from Equation~\eqref{eq:A_to_phi}. Figure~\ref{fig:supp:estimatedepsilon} shows all the individual estimations of $\epsilon_p-\epsilon_f$ together with the average value $\epsilon_p-\epsilon_f \approx 14.9 $~m$^{-1}$.

\begin{figure}
    \centering
    \includegraphics[scale=1] {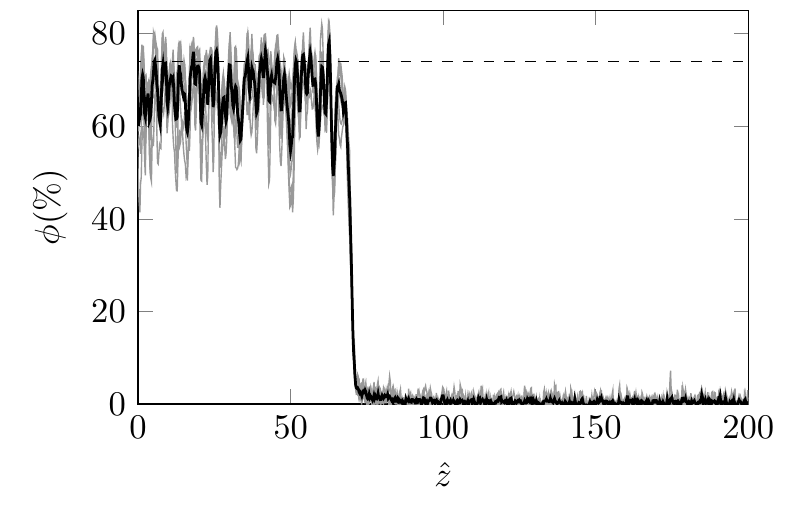}
    \caption{Particle volume fraction profiles near the outer wall of the Taylor-Couette geometry obtained with $\epsilon_p-\epsilon_f=14.9$~m$^{-1}$ for $\dot\gamma = 0$~s$^{-1}$. Gray lines correspond to a time-average of $\phi(r,z)$ at steady state for $r$ values corresponding to the 6 pixels closest to the outer cup on the raw images for both the left and right cross-sections. The thick black line is an average of the gray lines. The dashed line corresponds to $\phi_{\rm cfc} = 0.74$, which is the upper volume fraction limit for monodisperse spheres.}
    \label{fig:supp:normalization}
\end{figure}

 Such an extinction coefficient difference is compatible with an ordered layer at the outer edge of the experiment. We have plotted in Figure~\ref{fig:supp:normalization} the volume fraction profiles for the pixels corresponding to the outermost concentration profile zone (see Section~\ref{sec:extractprofiles} and Figure~\ref{fig:profiles_500s-1}) for the sediment at rest, using $\epsilon_p-\epsilon_f =14.9$. We notice that $\phi$ sometimes exceeds $0.74$, yet we have verified that the excursions above $\phi = 0.74$ never occur over more than one particle diameter, reflecting both on the ordering at the wall and the subparticle resolution of the X-ray apparatus. The average profile (Figure~\ref{fig:supp:normalization}, in black) shown in the main text only seldom exceeds this value, confirming our data validation scheme. 

Finally, we estimate the initial height $h_0$ to be used for the models in Section~\ref{sec:compar_conc} from the measured particle mass $M$ and by taking $\phi_m=0.6$ for the maximum volume fraction so that $h_0 =M/[\rho \pi (R_o^2-R_i^2) \phi_m] = 70.4 a$. In practice, since we observe discrepancies in the sediment height between the left and right cross-sections (see Figures~\ref{fig:meanfield} and~\ref{fig:allprofiles_leftright}), we correct $h_0$ for each cross-section based on the data in Figure~\ref{fig:supp:estimatedepsilon} and obtain $\hat{h}_0 = 72.8$ for the left CS and $68.6$ for the right CS.

\section{Spatiotemporal diagrams of resuspension experiments}
\label{sec:app:spatiotemp}

\begin{figure}
    \centering
    \includegraphics{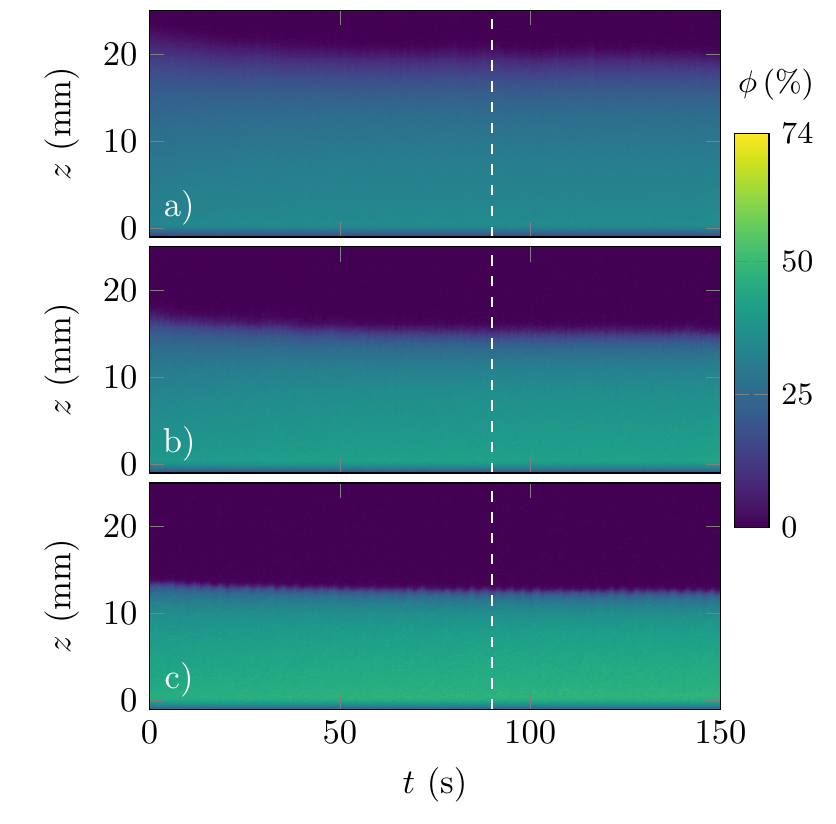}
    \caption{Spatiotemporal maps of the particle volume fraction for three different shear rates: a) $500$~s$^{-1}$; b) $100$~s$^{-1}$; c) $25$~s$^{-1}$. Each individual $\phi(r,z,t)$ at time $t$ is averaged over $r$ in order to provide a space-time representation in the $(z,t)$ plane. The mean volume fraction fields presented in Figure~\ref{fig:meanfield} are averaged over the last 120 images, i.e. from the white dashed line to the right end of the axis.}
    \label{fig:supp:spatiotemp}
\end{figure}

Figure~\ref{fig:supp:spatiotemp} displays particle volume fraction maps averaged over the gap width as a function of vertical position $z$ and time $t$ for three different applied shear rates. All three plots show that a steady profile is reached after $t \simeq 70$~s. Hence, the mean volume fraction fields shown in Figure~\ref{fig:meanfield}, where the temporal average was performed over $t=90$--150~s, do not include any transient regime and fully characterize viscous resuspension at the steady state.

\section{Normal viscosities for a different number of particles}
\label{sec:app:normalviscosity2g}

\begin{figure}
    \centering
    \includegraphics{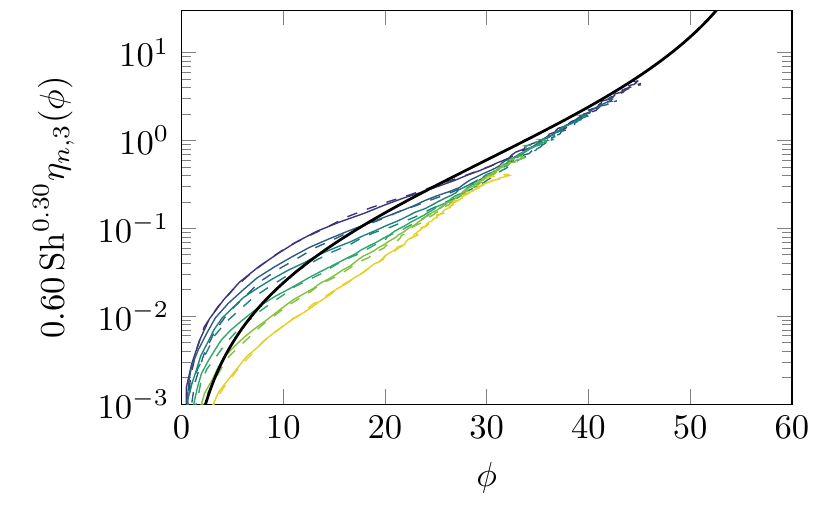}
    \caption{Rescaled normal viscosity $0.60 {\rm Sh}^{0.3} \eta_{n,3}$ as a function of volume fraction $\phi$ for experiments with a global volume fraction of 5\%, plotted following the results of Figure~\ref{fig:normalviscosities}. Colored lines depict experimental data: shear rates range from 25~s$^{-1}$ (blue) to 1000~s$^{-1}$ (yellow). The black solid line corresponds to the normal viscosity correlation proposed by Boyer~\emph{et al.}~\cite{Boyer2011}.}
    \label{fig:supp:normalviscosity_2g}
\end{figure}

Additional experiments have been conducted using only half of the particle number used in the main text, bringing the global volume fraction in the experiment down to $5\%$. Figure~\ref{fig:supp:normalviscosity_2g} shows the normal viscosities $\eta_{n,3}$ rescaled by the same factors as those used in Figure~\ref{fig:normalviscosities}. Discrepancies between shear rates are rather high for volume fractions below $\phi = 30\%$. This is to be expected since the resuspended sediment height is smaller, leading to larger parallax issues in the vertical direction. Nevertheless, their behavior at large volume fractions looks universal and matches the correlation proposed by \citet{Boyer2011} fairly well. This result confirms that resuspension does not depend on the total number of particles, and hence that the process is not strongly affected by boundary conditions.

\bibliography{biblio}

\begin{thebibliography}{37}%
\makeatletter
\providecommand \@ifxundefined [1]{%
 \@ifx{#1\undefined}
}%
\providecommand \@ifnum [1]{%
 \ifnum #1\expandafter \@firstoftwo
 \else \expandafter \@secondoftwo
 \fi
}%
\providecommand \@ifx [1]{%
 \ifx #1\expandafter \@firstoftwo
 \else \expandafter \@secondoftwo
 \fi
}%
\providecommand \natexlab [1]{#1}%
\providecommand \enquote  [1]{``#1''}%
\providecommand \bibnamefont  [1]{#1}%
\providecommand \bibfnamefont [1]{#1}%
\providecommand \citenamefont [1]{#1}%
\providecommand \href@noop [0]{\@secondoftwo}%
\providecommand \href [0]{\begingroup \@sanitize@url \@href}%
\providecommand \@href[1]{\@@startlink{#1}\@@href}%
\providecommand \@@href[1]{\endgroup#1\@@endlink}%
\providecommand \@sanitize@url [0]{\catcode `\\12\catcode `\$12\catcode
  `\&12\catcode `\#12\catcode `\^12\catcode `\_12\catcode `\%12\relax}%
\providecommand \@@startlink[1]{}%
\providecommand \@@endlink[0]{}%
\providecommand \url  [0]{\begingroup\@sanitize@url \@url }%
\providecommand \@url [1]{\endgroup\@href {#1}{\urlprefix }}%
\providecommand \urlprefix  [0]{URL }%
\providecommand \Eprint [0]{\href }%
\providecommand \doibase [0]{http://dx.doi.org/}%
\providecommand \selectlanguage [0]{\@gobble}%
\providecommand \bibinfo  [0]{\@secondoftwo}%
\providecommand \bibfield  [0]{\@secondoftwo}%
\providecommand \translation [1]{[#1]}%
\providecommand \BibitemOpen [0]{}%
\providecommand \bibitemStop [0]{}%
\providecommand \bibitemNoStop [0]{.\EOS\space}%
\providecommand \EOS [0]{\spacefactor3000\relax}%
\providecommand \BibitemShut  [1]{\csname bibitem#1\endcsname}%
\let\auto@bib@innerbib\@empty
\bibitem [{\citenamefont {Stokes}(1851)}]{Stokes1851}%
  \BibitemOpen
  \bibfield  {author} {\bibinfo {author} {\bibfnamefont {G.~G.}\ \bibnamefont
  {Stokes}},\ }\href@noop {} {\emph {\bibinfo {title} {On the effect of the
  internal friction of fluids on the motion of pendulums}}},\ Vol.~\bibinfo
  {volume} {9}\ (\bibinfo  {publisher} {Pitt Press Cambridge},\ \bibinfo {year}
  {1851})\BibitemShut {NoStop}%
\bibitem [{\citenamefont {Einstein}(1906)}]{Einstein1906}%
  \BibitemOpen
  \bibfield  {author} {\bibinfo {author} {\bibfnamefont {A.}~\bibnamefont
  {Einstein}},\ }\bibfield  {title} {\enquote {\bibinfo {title} {Eine neue
  {B}estimmung der {M}olek{\"u}ldimensionen},}\ }\href@noop {} {\bibfield
  {journal} {\bibinfo  {journal} {Ann. Phys.}\ }\textbf {\bibinfo {volume}
  {324}},\ \bibinfo {pages} {289--306} (\bibinfo {year} {1906})}\BibitemShut
  {NoStop}%
\bibitem [{\citenamefont {Nott}\ and\ \citenamefont {Brady}(1994)}]{Nott1994}%
  \BibitemOpen
  \bibfield  {author} {\bibinfo {author} {\bibfnamefont {P.~R.}\ \bibnamefont
  {Nott}}\ and\ \bibinfo {author} {\bibfnamefont {J.~F.}\ \bibnamefont
  {Brady}},\ }\bibfield  {title} {\enquote {\bibinfo {title} {Pressure-driven
  flow of suspensions: Simulation and theory},}\ }\href@noop {} {\bibfield
  {journal} {\bibinfo  {journal} {J. Fluid Mech.}\ }\textbf {\bibinfo {volume}
  {275}},\ \bibinfo {pages} {157--199} (\bibinfo {year} {1994})}\BibitemShut
  {NoStop}%
\bibitem [{\citenamefont {Morris}\ and\ \citenamefont
  {Boulay}(1999)}]{Morris1999}%
  \BibitemOpen
  \bibfield  {author} {\bibinfo {author} {\bibfnamefont {J.~F.}\ \bibnamefont
  {Morris}}\ and\ \bibinfo {author} {\bibfnamefont {F.}~\bibnamefont
  {Boulay}},\ }\bibfield  {title} {\enquote {\bibinfo {title} {Curvilinear
  flows of noncolloidal suspensions: The role of normal stresses},}\
  }\href@noop {} {\bibfield  {journal} {\bibinfo  {journal} {J. Rheol.}\
  }\textbf {\bibinfo {volume} {43}},\ \bibinfo {pages} {1213--1237} (\bibinfo
  {year} {1999})}\BibitemShut {NoStop}%
\bibitem [{\citenamefont {Lhuillier}(2009)}]{Lhuillier2009}%
  \BibitemOpen
  \bibfield  {author} {\bibinfo {author} {\bibfnamefont {D.}~\bibnamefont
  {Lhuillier}},\ }\bibfield  {title} {\enquote {\bibinfo {title} {Migration of
  rigid particles in non-{B}rownian viscous suspensions},}\ }\href@noop {}
  {\bibfield  {journal} {\bibinfo  {journal} {Phys. Fluids}\ }\textbf {\bibinfo
  {volume} {21}},\ \bibinfo {pages} {023302} (\bibinfo {year}
  {2009})}\BibitemShut {NoStop}%
\bibitem [{\citenamefont {Nott}, \citenamefont {Guazzelli},\ and\ \citenamefont
  {Pouliquen}(2011)}]{Nott2011}%
  \BibitemOpen
  \bibfield  {author} {\bibinfo {author} {\bibfnamefont {P.~R.}\ \bibnamefont
  {Nott}}, \bibinfo {author} {\bibfnamefont {{\'E}.}~\bibnamefont {Guazzelli}},
  \ and\ \bibinfo {author} {\bibfnamefont {O.}~\bibnamefont {Pouliquen}},\
  }\bibfield  {title} {\enquote {\bibinfo {title} {The suspension balance model
  revisited},}\ }\href@noop {} {\bibfield  {journal} {\bibinfo  {journal}
  {Phys. Fluids}\ }\textbf {\bibinfo {volume} {23}},\ \bibinfo {pages} {043304}
  (\bibinfo {year} {2011})}\BibitemShut {NoStop}%
\bibitem [{\citenamefont {Boyer}, \citenamefont {Guazzelli},\ and\
  \citenamefont {Pouliquen}(2011)}]{Boyer2011}%
  \BibitemOpen
  \bibfield  {author} {\bibinfo {author} {\bibfnamefont {F.}~\bibnamefont
  {Boyer}}, \bibinfo {author} {\bibfnamefont {{\'E}.}~\bibnamefont
  {Guazzelli}}, \ and\ \bibinfo {author} {\bibfnamefont {O.}~\bibnamefont
  {Pouliquen}},\ }\bibfield  {title} {\enquote {\bibinfo {title} {Unifying
  suspension and granular rheology},}\ }\href@noop {} {\bibfield  {journal}
  {\bibinfo  {journal} {Phys. Rev. Lett.}\ }\textbf {\bibinfo {volume} {107}},\
  \bibinfo {pages} {188301} (\bibinfo {year} {2011})}\BibitemShut {NoStop}%
\bibitem [{\citenamefont {Guazzelli}\ and\ \citenamefont
  {Pouliquen}(2018)}]{Guazzelli2018}%
  \BibitemOpen
  \bibfield  {author} {\bibinfo {author} {\bibfnamefont {{\'E}.}~\bibnamefont
  {Guazzelli}}\ and\ \bibinfo {author} {\bibfnamefont {O.}~\bibnamefont
  {Pouliquen}},\ }\bibfield  {title} {\enquote {\bibinfo {title} {Rheology of
  dense granular suspensions},}\ }\href@noop {} {\bibfield  {journal} {\bibinfo
   {journal} {J. Fluid Mech.}\ }\textbf {\bibinfo {volume} {852}} (\bibinfo
  {year} {2018})}\BibitemShut {NoStop}%
\bibitem [{\citenamefont {Guazzelli}\ and\ \citenamefont
  {Morris}(2012)}]{Guazzelli2012}%
  \BibitemOpen
  \bibfield  {author} {\bibinfo {author} {\bibfnamefont {{\'E}.}~\bibnamefont
  {Guazzelli}}\ and\ \bibinfo {author} {\bibfnamefont {J.~F.}\ \bibnamefont
  {Morris}},\ }\href@noop {} {\emph {\bibinfo {title} {A Physical Introduction
  to Suspension Dynamics}}}\ (\bibinfo  {publisher} {Cambridge University
  Press},\ \bibinfo {year} {2012})\BibitemShut {NoStop}%
\bibitem [{\citenamefont {Zarraga}, \citenamefont {Hill},\ and\ \citenamefont
  {{Leighton Jr.}}(2000)}]{Zarraga2000}%
  \BibitemOpen
  \bibfield  {author} {\bibinfo {author} {\bibfnamefont {I.~E.}\ \bibnamefont
  {Zarraga}}, \bibinfo {author} {\bibfnamefont {D.~A.}\ \bibnamefont {Hill}}, \
  and\ \bibinfo {author} {\bibfnamefont {D.~T.}\ \bibnamefont {{Leighton
  Jr.}}},\ }\bibfield  {title} {\enquote {\bibinfo {title} {The
  characterization of the total stress of concentrated suspensions of
  noncolloidal spheres in {N}ewtonian fluids},}\ }\href@noop {} {\bibfield
  {journal} {\bibinfo  {journal} {J. Rheol.}\ }\textbf {\bibinfo {volume}
  {44}},\ \bibinfo {pages} {185--220} (\bibinfo {year} {2000})}\BibitemShut
  {NoStop}%
\bibitem [{\citenamefont {Dbouk}, \citenamefont {Lobry},\ and\ \citenamefont
  {Lemaire}(2013)}]{Dbouk2013}%
  \BibitemOpen
  \bibfield  {author} {\bibinfo {author} {\bibfnamefont {T.}~\bibnamefont
  {Dbouk}}, \bibinfo {author} {\bibfnamefont {L.}~\bibnamefont {Lobry}}, \ and\
  \bibinfo {author} {\bibfnamefont {E.}~\bibnamefont {Lemaire}},\ }\bibfield
  {title} {\enquote {\bibinfo {title} {Normal stresses in concentrated
  non-{B}rownian suspensions},}\ }\href@noop {} {\bibfield  {journal} {\bibinfo
   {journal} {J. Fluid Mech.}\ }\textbf {\bibinfo {volume} {715}},\ \bibinfo
  {pages} {239--272} (\bibinfo {year} {2013})}\BibitemShut {NoStop}%
\bibitem [{\citenamefont {Chatt{\'e}}\ \emph {et~al.}(2018)\citenamefont
  {Chatt{\'e}}, \citenamefont {Comtet}, \citenamefont {Nigu{\`e}s},
  \citenamefont {Bocquet}, \citenamefont {Siria}, \citenamefont {Ducouret},
  \citenamefont {Lequeux}, \citenamefont {Lenoir}, \citenamefont {Ovarlez},\
  and\ \citenamefont {Colin}}]{Chatte2018}%
  \BibitemOpen
  \bibfield  {author} {\bibinfo {author} {\bibfnamefont {G.}~\bibnamefont
  {Chatt{\'e}}}, \bibinfo {author} {\bibfnamefont {J.}~\bibnamefont {Comtet}},
  \bibinfo {author} {\bibfnamefont {A.}~\bibnamefont {Nigu{\`e}s}}, \bibinfo
  {author} {\bibfnamefont {L.}~\bibnamefont {Bocquet}}, \bibinfo {author}
  {\bibfnamefont {A.}~\bibnamefont {Siria}}, \bibinfo {author} {\bibfnamefont
  {G.}~\bibnamefont {Ducouret}}, \bibinfo {author} {\bibfnamefont
  {F.}~\bibnamefont {Lequeux}}, \bibinfo {author} {\bibfnamefont
  {N.}~\bibnamefont {Lenoir}}, \bibinfo {author} {\bibfnamefont
  {G.}~\bibnamefont {Ovarlez}}, \ and\ \bibinfo {author} {\bibfnamefont
  {A.}~\bibnamefont {Colin}},\ }\bibfield  {title} {\enquote {\bibinfo {title}
  {Shear thinning in non-{B}rownian suspensions},}\ }\href@noop {} {\bibfield
  {journal} {\bibinfo  {journal} {Soft Matt.}\ }\textbf {\bibinfo {volume}
  {14}},\ \bibinfo {pages} {879--893} (\bibinfo {year} {2018})}\BibitemShut
  {NoStop}%
\bibitem [{\citenamefont {Tanner}\ \emph {et~al.}(2018)\citenamefont {Tanner},
  \citenamefont {Ness}, \citenamefont {Mahmud}, \citenamefont {Dai},\ and\
  \citenamefont {Moon}}]{Tanner2018}%
  \BibitemOpen
  \bibfield  {author} {\bibinfo {author} {\bibfnamefont {R.~I.}\ \bibnamefont
  {Tanner}}, \bibinfo {author} {\bibfnamefont {C.}~\bibnamefont {Ness}},
  \bibinfo {author} {\bibfnamefont {A.}~\bibnamefont {Mahmud}}, \bibinfo
  {author} {\bibfnamefont {S.}~\bibnamefont {Dai}}, \ and\ \bibinfo {author}
  {\bibfnamefont {J.}~\bibnamefont {Moon}},\ }\bibfield  {title} {\enquote
  {\bibinfo {title} {A bootstrap mechanism for non-colloidal suspension
  viscosity},}\ }\href@noop {} {\bibfield  {journal} {\bibinfo  {journal}
  {Rheol. Acta}\ }\textbf {\bibinfo {volume} {57}},\ \bibinfo {pages}
  {635--643} (\bibinfo {year} {2018})}\BibitemShut {NoStop}%
\bibitem [{\citenamefont {Lobry}\ \emph {et~al.}(2019)\citenamefont {Lobry},
  \citenamefont {Lemaire}, \citenamefont {Blanc}, \citenamefont {Gallier},\
  and\ \citenamefont {Peters}}]{Lobry2019}%
  \BibitemOpen
  \bibfield  {author} {\bibinfo {author} {\bibfnamefont {L.}~\bibnamefont
  {Lobry}}, \bibinfo {author} {\bibfnamefont {E.}~\bibnamefont {Lemaire}},
  \bibinfo {author} {\bibfnamefont {F.}~\bibnamefont {Blanc}}, \bibinfo
  {author} {\bibfnamefont {S.}~\bibnamefont {Gallier}}, \ and\ \bibinfo
  {author} {\bibfnamefont {F.}~\bibnamefont {Peters}},\ }\bibfield  {title}
  {\enquote {\bibinfo {title} {Shear thinning in non-{B}rownian suspensions
  explained by variable friction between particles},}\ }\href@noop {}
  {\bibfield  {journal} {\bibinfo  {journal} {J. Fluid Mech.}\ }\textbf
  {\bibinfo {volume} {860}},\ \bibinfo {pages} {682--710} (\bibinfo {year}
  {2019})}\BibitemShut {NoStop}%
\bibitem [{\citenamefont {Gamonpilas}, \citenamefont {Morris},\ and\
  \citenamefont {Denn}(2016)}]{Gamonpilas2016}%
  \BibitemOpen
  \bibfield  {author} {\bibinfo {author} {\bibfnamefont {C.}~\bibnamefont
  {Gamonpilas}}, \bibinfo {author} {\bibfnamefont {J.~F.}\ \bibnamefont
  {Morris}}, \ and\ \bibinfo {author} {\bibfnamefont {M.~M.}\ \bibnamefont
  {Denn}},\ }\bibfield  {title} {\enquote {\bibinfo {title} {Shear and normal
  stress measurements in non-{B}rownian monodisperse and bidisperse
  suspensions},}\ }\href@noop {} {\bibfield  {journal} {\bibinfo  {journal} {J.
  Rheol.}\ }\textbf {\bibinfo {volume} {60}},\ \bibinfo {pages} {289--296}
  (\bibinfo {year} {2016})}\BibitemShut {NoStop}%
\bibitem [{\citenamefont {Couturier}\ \emph {et~al.}(2011)\citenamefont
  {Couturier}, \citenamefont {Boyer}, \citenamefont {Pouliquen},\ and\
  \citenamefont {Guazzelli}}]{Couturier2011}%
  \BibitemOpen
  \bibfield  {author} {\bibinfo {author} {\bibfnamefont {{\'E}.}~\bibnamefont
  {Couturier}}, \bibinfo {author} {\bibfnamefont {F.}~\bibnamefont {Boyer}},
  \bibinfo {author} {\bibfnamefont {O.}~\bibnamefont {Pouliquen}}, \ and\
  \bibinfo {author} {\bibfnamefont {{\'E}.}~\bibnamefont {Guazzelli}},\
  }\bibfield  {title} {\enquote {\bibinfo {title} {Suspensions in a tilted
  trough: Second normal stress difference},}\ }\href@noop {} {\bibfield
  {journal} {\bibinfo  {journal} {J. Fluid Mech.}\ }\textbf {\bibinfo {volume}
  {686}},\ \bibinfo {pages} {26--39} (\bibinfo {year} {2011})}\BibitemShut
  {NoStop}%
\bibitem [{\citenamefont {{Leighton Jr.}}\ and\ \citenamefont
  {Acrivos}(1986)}]{Leighton1986}%
  \BibitemOpen
  \bibfield  {author} {\bibinfo {author} {\bibfnamefont {D.~T.}\ \bibnamefont
  {{Leighton Jr.}}}\ and\ \bibinfo {author} {\bibfnamefont {A.}~\bibnamefont
  {Acrivos}},\ }\bibfield  {title} {\enquote {\bibinfo {title} {Viscous
  resuspension},}\ }\href@noop {} {\bibfield  {journal} {\bibinfo  {journal}
  {Chem. Eng. Sci.}\ }\textbf {\bibinfo {volume} {41}},\ \bibinfo {pages}
  {1377--1384} (\bibinfo {year} {1986})}\BibitemShut {NoStop}%
\bibitem [{\citenamefont {Acrivos}, \citenamefont {Mauri},\ and\ \citenamefont
  {Fan}(1993)}]{Acrivos1993}%
  \BibitemOpen
  \bibfield  {author} {\bibinfo {author} {\bibfnamefont {A.}~\bibnamefont
  {Acrivos}}, \bibinfo {author} {\bibfnamefont {R.}~\bibnamefont {Mauri}}, \
  and\ \bibinfo {author} {\bibfnamefont {X.}~\bibnamefont {Fan}},\ }\bibfield
  {title} {\enquote {\bibinfo {title} {Shear-induced resuspension in a
  {C}ouette device},}\ }\href@noop {} {\bibfield  {journal} {\bibinfo
  {journal} {Int. J. of Multiph. Flow}\ }\textbf {\bibinfo {volume} {19}},\
  \bibinfo {pages} {797--802} (\bibinfo {year} {1993})}\BibitemShut {NoStop}%
\bibitem [{\citenamefont {Deboeuf}\ \emph {et~al.}(2018)\citenamefont
  {Deboeuf}, \citenamefont {Lenoir}, \citenamefont {Hautemayou}, \citenamefont
  {Bornert}, \citenamefont {Blanc},\ and\ \citenamefont
  {Ovarlez}}]{Deboeuf2018}%
  \BibitemOpen
  \bibfield  {author} {\bibinfo {author} {\bibfnamefont {S.}~\bibnamefont
  {Deboeuf}}, \bibinfo {author} {\bibfnamefont {N.}~\bibnamefont {Lenoir}},
  \bibinfo {author} {\bibfnamefont {D.}~\bibnamefont {Hautemayou}}, \bibinfo
  {author} {\bibfnamefont {M.}~\bibnamefont {Bornert}}, \bibinfo {author}
  {\bibfnamefont {F.}~\bibnamefont {Blanc}}, \ and\ \bibinfo {author}
  {\bibfnamefont {G.}~\bibnamefont {Ovarlez}},\ }\bibfield  {title} {\enquote
  {\bibinfo {title} {Imaging non-{B}rownian particle suspensions with {X}-ray
  tomography: Application to the microstructure of {N}ewtonian and viscoplastic
  suspensions},}\ }\href@noop {} {\bibfield  {journal} {\bibinfo  {journal} {J.
  Rheol.}\ }\textbf {\bibinfo {volume} {62}},\ \bibinfo {pages} {643--663}
  (\bibinfo {year} {2018})}\BibitemShut {NoStop}%
\bibitem [{\citenamefont {Gholami}\ \emph {et~al.}(2018)\citenamefont
  {Gholami}, \citenamefont {Rashedi}, \citenamefont {Lenoir}, \citenamefont
  {Hautemayou}, \citenamefont {Ovarlez},\ and\ \citenamefont
  {Hormozi}}]{Gholami2018}%
  \BibitemOpen
  \bibfield  {author} {\bibinfo {author} {\bibfnamefont {M.}~\bibnamefont
  {Gholami}}, \bibinfo {author} {\bibfnamefont {A.}~\bibnamefont {Rashedi}},
  \bibinfo {author} {\bibfnamefont {N.}~\bibnamefont {Lenoir}}, \bibinfo
  {author} {\bibfnamefont {D.}~\bibnamefont {Hautemayou}}, \bibinfo {author}
  {\bibfnamefont {G.}~\bibnamefont {Ovarlez}}, \ and\ \bibinfo {author}
  {\bibfnamefont {S.}~\bibnamefont {Hormozi}},\ }\bibfield  {title} {\enquote
  {\bibinfo {title} {Time-resolved 2d concentration maps in flowing suspensions
  using {X}-ray},}\ }\href@noop {} {\bibfield  {journal} {\bibinfo  {journal}
  {J. Rheol.}\ }\textbf {\bibinfo {volume} {62}},\ \bibinfo {pages} {955--974}
  (\bibinfo {year} {2018})}\BibitemShut {NoStop}%
\bibitem [{\citenamefont {Bacri}\ \emph {et~al.}(1986)\citenamefont {Bacri},
  \citenamefont {Frenois}, \citenamefont {Hoyos}, \citenamefont {Perzynski},
  \citenamefont {Rakotomalala},\ and\ \citenamefont {Salin}}]{Bacri1986}%
  \BibitemOpen
  \bibfield  {author} {\bibinfo {author} {\bibfnamefont {J.-C.}\ \bibnamefont
  {Bacri}}, \bibinfo {author} {\bibfnamefont {C.}~\bibnamefont {Frenois}},
  \bibinfo {author} {\bibfnamefont {M.}~\bibnamefont {Hoyos}}, \bibinfo
  {author} {\bibfnamefont {R.}~\bibnamefont {Perzynski}}, \bibinfo {author}
  {\bibfnamefont {N.}~\bibnamefont {Rakotomalala}}, \ and\ \bibinfo {author}
  {\bibfnamefont {D.}~\bibnamefont {Salin}},\ }\bibfield  {title} {\enquote
  {\bibinfo {title} {Acoustic study of suspension sedimentation},}\ }\href@noop
  {} {\bibfield  {journal} {\bibinfo  {journal} {Europhys. Lett.}\ }\textbf
  {\bibinfo {volume} {2}},\ \bibinfo {pages} {123} (\bibinfo {year}
  {1986})}\BibitemShut {NoStop}%
\bibitem [{\citenamefont {Onoda}\ and\ \citenamefont
  {Liniger}(1990)}]{Onoda1990}%
  \BibitemOpen
  \bibfield  {author} {\bibinfo {author} {\bibfnamefont {G.~Y.}\ \bibnamefont
  {Onoda}}\ and\ \bibinfo {author} {\bibfnamefont {E.~G.}\ \bibnamefont
  {Liniger}},\ }\bibfield  {title} {\enquote {\bibinfo {title} {Random loose
  packings of uniform spheres and the dilatancy onset},}\ }\href {\doibase
  10.1103/PhysRevLett.64.2727} {\bibfield  {journal} {\bibinfo  {journal}
  {Phys. Rev. Lett.}\ }\textbf {\bibinfo {volume} {64}},\ \bibinfo {pages}
  {2727--2730} (\bibinfo {year} {1990})}\BibitemShut {NoStop}%
\bibitem [{\citenamefont {Dong}\ \emph {et~al.}(2006)\citenamefont {Dong},
  \citenamefont {Yang}, \citenamefont {Zou},\ and\ \citenamefont
  {Yu}}]{Dong2006}%
  \BibitemOpen
  \bibfield  {author} {\bibinfo {author} {\bibfnamefont {K.~J.}\ \bibnamefont
  {Dong}}, \bibinfo {author} {\bibfnamefont {R.~Y.}\ \bibnamefont {Yang}},
  \bibinfo {author} {\bibfnamefont {R.~P.}\ \bibnamefont {Zou}}, \ and\
  \bibinfo {author} {\bibfnamefont {A.~B.}\ \bibnamefont {Yu}},\ }\bibfield
  {title} {\enquote {\bibinfo {title} {Role of interparticle forces in the
  formation of random loose packing},}\ }\href {\doibase
  10.1103/PhysRevLett.96.145505} {\bibfield  {journal} {\bibinfo  {journal}
  {Phys. Rev. Lett.}\ }\textbf {\bibinfo {volume} {96}},\ \bibinfo {pages}
  {145505} (\bibinfo {year} {2006})}\BibitemShut {NoStop}%
\bibitem [{\citenamefont {Jerkins}\ \emph {et~al.}(2008)\citenamefont
  {Jerkins}, \citenamefont {Schr{\"{o}}ter}, \citenamefont {Swinney},
  \citenamefont {Senden}, \citenamefont {Saadatfar},\ and\ \citenamefont
  {Aste}}]{Jerkins2008}%
  \BibitemOpen
  \bibfield  {author} {\bibinfo {author} {\bibfnamefont {M.}~\bibnamefont
  {Jerkins}}, \bibinfo {author} {\bibfnamefont {M.}~\bibnamefont
  {Schr{\"{o}}ter}}, \bibinfo {author} {\bibfnamefont {H.~L.}\ \bibnamefont
  {Swinney}}, \bibinfo {author} {\bibfnamefont {T.~J.}\ \bibnamefont {Senden}},
  \bibinfo {author} {\bibfnamefont {M.}~\bibnamefont {Saadatfar}}, \ and\
  \bibinfo {author} {\bibfnamefont {T.}~\bibnamefont {Aste}},\ }\bibfield
  {title} {\enquote {\bibinfo {title} {Onset of mechanical stability in random
  packings of frictional spheres},}\ }\href {\doibase
  10.1103/physrevlett.101.018301} {\bibfield  {journal} {\bibinfo  {journal}
  {Phys. Rev. Lett.}\ }\textbf {\bibinfo {volume} {101}},\ \bibinfo {pages}
  {018301} (\bibinfo {year} {2008})}\BibitemShut {NoStop}%
\bibitem [{\citenamefont {Ovarlez}, \citenamefont {Bertrand},\ and\
  \citenamefont {Rodts}(2006)}]{Ovarlez2006}%
  \BibitemOpen
  \bibfield  {author} {\bibinfo {author} {\bibfnamefont {G.}~\bibnamefont
  {Ovarlez}}, \bibinfo {author} {\bibfnamefont {F.}~\bibnamefont {Bertrand}}, \
  and\ \bibinfo {author} {\bibfnamefont {S.}~\bibnamefont {Rodts}},\ }\bibfield
   {title} {\enquote {\bibinfo {title} {Local determination of the constitutive
  law of a dense suspension of noncolloidal particles through magnetic
  resonance imaging},}\ }\href@noop {} {\bibfield  {journal} {\bibinfo
  {journal} {J. Rheol.}\ }\textbf {\bibinfo {volume} {50}},\ \bibinfo {pages}
  {259--292} (\bibinfo {year} {2006})}\BibitemShut {NoStop}%
\bibitem [{\citenamefont {Mari}\ \emph {et~al.}(2015)\citenamefont {Mari},
  \citenamefont {Seto}, \citenamefont {Morris},\ and\ \citenamefont
  {Denn}}]{Mari2015}%
  \BibitemOpen
  \bibfield  {author} {\bibinfo {author} {\bibfnamefont {R.}~\bibnamefont
  {Mari}}, \bibinfo {author} {\bibfnamefont {R.}~\bibnamefont {Seto}}, \bibinfo
  {author} {\bibfnamefont {J.~F.}\ \bibnamefont {Morris}}, \ and\ \bibinfo
  {author} {\bibfnamefont {M.~M.}\ \bibnamefont {Denn}},\ }\bibfield  {title}
  {\enquote {\bibinfo {title} {Nonmonotonic flow curves of shear thickening
  suspensions},}\ }\href@noop {} {\bibfield  {journal} {\bibinfo  {journal}
  {Phys. Rev. E}\ }\textbf {\bibinfo {volume} {91}},\ \bibinfo {pages} {052302}
  (\bibinfo {year} {2015})}\BibitemShut {NoStop}%
\bibitem [{\citenamefont {Boyer}, \citenamefont {Pouliquen},\ and\
  \citenamefont {Guazzelli}(2011)}]{Boyer2011a}%
  \BibitemOpen
  \bibfield  {author} {\bibinfo {author} {\bibfnamefont {F.}~\bibnamefont
  {Boyer}}, \bibinfo {author} {\bibfnamefont {O.}~\bibnamefont {Pouliquen}}, \
  and\ \bibinfo {author} {\bibfnamefont {{\'E}.}~\bibnamefont {Guazzelli}},\
  }\bibfield  {title} {\enquote {\bibinfo {title} {Dense suspensions in
  rotating-rod flows: normal stresses and particle migration},}\ }\href
  {\doibase 10.1017/jfm.2011.272} {\bibfield  {journal} {\bibinfo  {journal}
  {J. Fluid Mech.}\ }\textbf {\bibinfo {volume} {686}},\ \bibinfo {pages}
  {5–25} (\bibinfo {year} {2011})}\BibitemShut {NoStop}%
\bibitem [{\citenamefont {Wang}, \citenamefont {Mauri},\ and\ \citenamefont
  {Acrivos}(1998)}]{Wang1998}%
  \BibitemOpen
  \bibfield  {author} {\bibinfo {author} {\bibfnamefont {Y.}~\bibnamefont
  {Wang}}, \bibinfo {author} {\bibfnamefont {R.}~\bibnamefont {Mauri}}, \ and\
  \bibinfo {author} {\bibfnamefont {A.}~\bibnamefont {Acrivos}},\ }\bibfield
  {title} {\enquote {\bibinfo {title} {Transverse shear-induced gradient
  diffusion in a dilute suspension of spheres},}\ }\href@noop {} {\bibfield
  {journal} {\bibinfo  {journal} {Journal of Fluid Mechanics}\ }\textbf
  {\bibinfo {volume} {357}},\ \bibinfo {pages} {279--287} (\bibinfo {year}
  {1998})}\BibitemShut {NoStop}%
\bibitem [{\citenamefont {Gadala-Maria}\ and\ \citenamefont
  {Acrivos}(1980)}]{Gadala-Maria1980}%
  \BibitemOpen
  \bibfield  {author} {\bibinfo {author} {\bibfnamefont {F.}~\bibnamefont
  {Gadala-Maria}}\ and\ \bibinfo {author} {\bibfnamefont {A.}~\bibnamefont
  {Acrivos}},\ }\bibfield  {title} {\enquote {\bibinfo {title} {Shear-induced
  structure in a concentrated suspension of solid spheres},}\ }\href@noop {}
  {\bibfield  {journal} {\bibinfo  {journal} {J. Rheol.}\ }\textbf {\bibinfo
  {volume} {24}},\ \bibinfo {pages} {799--814} (\bibinfo {year}
  {1980})}\BibitemShut {NoStop}%
\bibitem [{\citenamefont {Blanc}, \citenamefont {Peters},\ and\ \citenamefont
  {Lemaire}(2011)}]{blanc2011local}%
  \BibitemOpen
  \bibfield  {author} {\bibinfo {author} {\bibfnamefont {F.}~\bibnamefont
  {Blanc}}, \bibinfo {author} {\bibfnamefont {F.}~\bibnamefont {Peters}}, \
  and\ \bibinfo {author} {\bibfnamefont {E.}~\bibnamefont {Lemaire}},\
  }\bibfield  {title} {\enquote {\bibinfo {title} {Local transient rheological
  behavior of concentrated suspensions},}\ }\href@noop {} {\bibfield  {journal}
  {\bibinfo  {journal} {J. Rheol.}\ }\textbf {\bibinfo {volume} {55}},\
  \bibinfo {pages} {835--854} (\bibinfo {year} {2011})}\BibitemShut {NoStop}%
\bibitem [{\citenamefont {Lin}\ \emph {et~al.}(2015)\citenamefont {Lin},
  \citenamefont {Guy}, \citenamefont {Hermes}, \citenamefont {Ness},
  \citenamefont {Sun}, \citenamefont {Poon},\ and\ \citenamefont
  {Cohen}}]{Lin2015}%
  \BibitemOpen
  \bibfield  {author} {\bibinfo {author} {\bibfnamefont {N.~Y.}\ \bibnamefont
  {Lin}}, \bibinfo {author} {\bibfnamefont {B.~M.}\ \bibnamefont {Guy}},
  \bibinfo {author} {\bibfnamefont {M.}~\bibnamefont {Hermes}}, \bibinfo
  {author} {\bibfnamefont {C.}~\bibnamefont {Ness}}, \bibinfo {author}
  {\bibfnamefont {J.}~\bibnamefont {Sun}}, \bibinfo {author} {\bibfnamefont
  {W.~C.}\ \bibnamefont {Poon}}, \ and\ \bibinfo {author} {\bibfnamefont
  {I.}~\bibnamefont {Cohen}},\ }\bibfield  {title} {\enquote {\bibinfo {title}
  {Hydrodynamic and contact contributions to continuous shear thickening in
  colloidal suspensions},}\ }\href@noop {} {\bibfield  {journal} {\bibinfo
  {journal} {Phys. Rev. Lett.}\ }\textbf {\bibinfo {volume} {115}},\ \bibinfo
  {pages} {228304} (\bibinfo {year} {2015})}\BibitemShut {NoStop}%
\bibitem [{\citenamefont {Peters}\ \emph {et~al.}(2016)\citenamefont {Peters},
  \citenamefont {Ghigliotti}, \citenamefont {Gallier}, \citenamefont {Blanc},
  \citenamefont {Lemaire},\ and\ \citenamefont {Lobry}}]{Peters2016}%
  \BibitemOpen
  \bibfield  {author} {\bibinfo {author} {\bibfnamefont {F.}~\bibnamefont
  {Peters}}, \bibinfo {author} {\bibfnamefont {G.}~\bibnamefont {Ghigliotti}},
  \bibinfo {author} {\bibfnamefont {S.}~\bibnamefont {Gallier}}, \bibinfo
  {author} {\bibfnamefont {F.}~\bibnamefont {Blanc}}, \bibinfo {author}
  {\bibfnamefont {E.}~\bibnamefont {Lemaire}}, \ and\ \bibinfo {author}
  {\bibfnamefont {L.}~\bibnamefont {Lobry}},\ }\bibfield  {title} {\enquote
  {\bibinfo {title} {Rheology of non-{B}rownian suspensions of rough frictional
  particles under shear reversal: A numerical study},}\ }\href@noop {}
  {\bibfield  {journal} {\bibinfo  {journal} {J. Rheol.}\ }\textbf {\bibinfo
  {volume} {60}},\ \bibinfo {pages} {715--732} (\bibinfo {year}
  {2016})}\BibitemShut {NoStop}%
\bibitem [{\citenamefont {Maron}\ and\ \citenamefont
  {Pierce}(1956)}]{Maron1956}%
  \BibitemOpen
  \bibfield  {author} {\bibinfo {author} {\bibfnamefont {S.~H.}\ \bibnamefont
  {Maron}}\ and\ \bibinfo {author} {\bibfnamefont {P.~E.}\ \bibnamefont
  {Pierce}},\ }\bibfield  {title} {\enquote {\bibinfo {title} {Application of
  {R}ee-{E}yring generalized flow theory to suspensions of spherical
  particles},}\ }\href@noop {} {\bibfield  {journal} {\bibinfo  {journal}
  {Journal of colloid science}\ }\textbf {\bibinfo {volume} {11}},\ \bibinfo
  {pages} {80--95} (\bibinfo {year} {1956})}\BibitemShut {NoStop}%
\bibitem [{\citenamefont {Krieger}\ and\ \citenamefont
  {Dougherty}(1959)}]{Krieger1959}%
  \BibitemOpen
  \bibfield  {author} {\bibinfo {author} {\bibfnamefont {I.~M.}\ \bibnamefont
  {Krieger}}\ and\ \bibinfo {author} {\bibfnamefont {T.~J.}\ \bibnamefont
  {Dougherty}},\ }\bibfield  {title} {\enquote {\bibinfo {title} {A mechanism
  for non-{N}ewtonian flow in suspensions of rigid spheres},}\ }\href@noop {}
  {\bibfield  {journal} {\bibinfo  {journal} {Transactions of the Society of
  Rheology}\ }\textbf {\bibinfo {volume} {3}},\ \bibinfo {pages} {137--152}
  (\bibinfo {year} {1959})}\BibitemShut {NoStop}%
\bibitem [{\citenamefont {Jana}, \citenamefont {Kapoor},\ and\ \citenamefont
  {Acrivos}(1995)}]{Jana1995}%
  \BibitemOpen
  \bibfield  {author} {\bibinfo {author} {\bibfnamefont {S.}~\bibnamefont
  {Jana}}, \bibinfo {author} {\bibfnamefont {B.}~\bibnamefont {Kapoor}}, \ and\
  \bibinfo {author} {\bibfnamefont {A.}~\bibnamefont {Acrivos}},\ }\bibfield
  {title} {\enquote {\bibinfo {title} {Apparent wall slip velocity coefficients
  in concentrated suspensions of noncolloidal particles},}\ }\href@noop {}
  {\bibfield  {journal} {\bibinfo  {journal} {J. Rheol.}\ }\textbf {\bibinfo
  {volume} {39}},\ \bibinfo {pages} {1123--1132} (\bibinfo {year}
  {1995})}\BibitemShut {NoStop}%
\bibitem [{\citenamefont {Andreotti}, \citenamefont {Forterre},\ and\
  \citenamefont {Pouliquen}(2013)}]{Andreotti2013}%
  \BibitemOpen
  \bibfield  {author} {\bibinfo {author} {\bibfnamefont {B.}~\bibnamefont
  {Andreotti}}, \bibinfo {author} {\bibfnamefont {Y.}~\bibnamefont {Forterre}},
  \ and\ \bibinfo {author} {\bibfnamefont {O.}~\bibnamefont {Pouliquen}},\
  }\href@noop {} {\emph {\bibinfo {title} {Granular Media: Between Fluid and
  Solid}}}\ (\bibinfo  {publisher} {Cambridge University Press},\ \bibinfo
  {year} {2013})\BibitemShut {NoStop}%
\bibitem [{\citenamefont {D'Ambrosio}\ \emph {et~al.}(2019)\citenamefont
  {D'Ambrosio}, \citenamefont {Blanc}, \citenamefont {Peters}, \citenamefont
  {Lobry},\ and\ \citenamefont {Lemaire}}]{DAmbrosio2019}%
  \BibitemOpen
  \bibfield  {author} {\bibinfo {author} {\bibfnamefont {E.}~\bibnamefont
  {D'Ambrosio}}, \bibinfo {author} {\bibfnamefont {F.}~\bibnamefont {Blanc}},
  \bibinfo {author} {\bibfnamefont {F.}~\bibnamefont {Peters}}, \bibinfo
  {author} {\bibfnamefont {L.}~\bibnamefont {Lobry}}, \ and\ \bibinfo {author}
  {\bibfnamefont {E.}~\bibnamefont {Lemaire}},\ }\href@noop {} {} (\bibinfo
  {year} {2019}),\ \bibinfo {note} {in preparation}\BibitemShut {NoStop}%
\end{thebibliography}%
    
\end{document}